\documentclass[aps,prd,nofootinbib,superscriptaddress,twocolumn]{revtex4-2}
\usepackage{amsmath}
\usepackage{amsthm}
\usepackage{amsfonts}
\usepackage{natbib}
\usepackage{graphicx}
\usepackage[usenames,dvipsnames]{color}
\usepackage{orcidlink}

\def\lsim{~\rlap{$<$}{\lower 1.0ex\hbox{$\sim$}}}
\def\bsim{~\rlap{$>$}{\lower 1.0ex\hbox{$\sim$}}}

\def\apjl{Astrophys.\ J.\ Lett.}
\def\mnras{Mon.\ Not.\ R.\ Astron.\ Soc.}

\def\araa{Ann. Rev. Astron. Astrophys.}
\def\aap{Astron.\ Astrophys.}
\def\apj{Astrophys.\ J.}

\def\prd{Phys.\ Rev.\ D}

\def\jcap{{JCAP\ }}

\newcommand{\lmax}{\ell_\text{max}}
\newcommand{\jmax}{j_\text{max}}

\def\ln{{\rm ln}}

\def\mathbi#1{\textbf{\em #1}}

\def\rvh{\mathrm{\hat{\bf{r}}}}
\def\xvh{\mathrm{\hat{\bf{x}}}}
\def\yvh{\mathrm{\hat{\bf{y}}}}

\def\vk{\mathbi{k}}

\def\vr{\mathbi{r}}

\def\vx{\mathbi{x}}

\def\vvf{\mathbi{F}}

\def\grad{\boldsymbol{\nabla}}

\def\mach{\mathcal{M}}
\def\delrho{\delta\rho}
\def\gjlm{g^{(j)}_{\ell m}}
\def\fjl{f_{j \ell}}
\def\phiw{\phi}
\def\dphiw{\dot\phi}
\def\delw{\delta\rho}
\def\velw{\boldsymbol{v}}
\def\rapo{r_\text{apo}}
\def\vpmax{v_\text{peri}}

\allowdisplaybreaks

\begin{document}

\title{Harmonic-decomposition approach to dynamical friction for eccentric orbits}

\author{Gali Eytan\,\orcidlink{0009-0009-0366-7103}}
\email{gali.eytan@campus.technion.ac.il}
\affiliation{Physics Department, Technion -- Israel Institute of Technology, Haifa 3200003, Israel}

\author{Vincent Desjacques\,\orcidlink{0000-0003-2062-8172}}
\email{dvince@physics.technion.ac.il}
\affiliation{Physics Department, Technion -- Israel Institute of Technology, Haifa 3200003, Israel}

\author{Yonadav Barry Ginat\,\orcidlink{0000-0003-1992-1910}}
\email{yb.ginat@physics.ox.ac.uk}%
\affiliation{Rudolf Peierls Centre for Theoretical Physics, University of Oxford, Parks Road, Oxford, OX1 3PU, United Kingdom}%
\affiliation{New College, Holywell Street, Oxford, OX1 3BN, United Kingdom}%

\date{\today}

\begin{abstract}
  Compact objects evolving in an astrophysical environment experience a gravitational drag force known as dynamical friction. 
  We present a multipole–frequency decomposition to evaluate the orbit-averaged energy and angular momentum dissipation experienced by point masses on periodic orbits 
  within a homogeneous, fluid-like background. Our focus is on eccentric Keplerian trajectories. 
  Although our approach is currently restricted to linear response theory, it is fully consistent within that framework. 
  We validate our theoretical expressions for the specific case of an ideal fluid, using semi-numerical simulations of the linear response acoustic wake.
  We demonstrate that, for a finite time perturbation switched on at $t=0$, a steady dissipation state is reached after a time bounded by twice the sound crossing time 
  of the apocentre distance. 
  We apply our results to model the secular evolution of compact eccentric binaries in a gaseous medium, assuming low-density conditions where the orbital elements 
  evolve adiabatically. 
  For unequal-mass systems with moderate initial eccentricity, the late-time eccentricity growth is significantly delayed compared to the equal-mass case, due to the
  binary components becoming transonic at different times along their orbital trajectory.
  Our approach offers a computationally efficient alternative to full simulations of the linear response wake.
\end{abstract}

\maketitle

\section{Introduction}
\label{sec:intro}

Astrophysical bodies across a wide range of masses, from stellar binaries to galaxy clusters, merge through the loss of orbital energy. 
In non-vacuum environments such as the interstellar medium or a dark matter halo, this energy loss can be influenced by a drag force arising from the medium’s response 
to the gravitational disturbance caused by the merging objects. 
This dissipative process known as dynamical friction (DF) is a fundamental ingredient in the hierarchical growth of cosmic structures 
\citep{chandrasekhar:1943,tremaine/etal:1975,paczynski:1976,murai/fujimoto:1980,binney/tremaine:1987,kauffmann/etal:1993,ibata/lewis:1998,somerville/primack:1999,vandenbosch/etal:1999,ostriker:1999,cole/etal:2000,goldreich/etal:2004,croton/etal:2006,boylankolchin/etal:2008,mo/etal:2010}.
The study of this effect was pioneered in \cite{chandrasekhar:1943}, who calculated the DF force produced by the density wake trailing 
a point mass moving uniformly through an infinite, homogeneous, and collisionless medium.
Since then, extensive work has been devoted to understanding the physics of dynamical friction along various perturber trajectories and in different backgrounds, 
which include stellar distributions \citep{kalnajs:1971,kalnajs:1972,lyndenbell/kalnajs:1972,tremaine/weinberg:1984,palmer/papaloizou:1985,weinberg:1986,bekenstein/maoz:1992,nelson/tremaine:1999,banik/vandenbosch:2021,kaur/stone:2022,ginat/etal:2023,kipper/etal:2023}, 
gaseous environments \citep{dokuchaev:1964,ruderman/spiegel:1971,rephaeli/salpeter:1980,just/kegel:1990,ostriker:1999,salcedo/brandenburg:2001,kim/kim:2007,lee/stahler:2011,vicente/etal:2019,salcedo:2019,desjacques/etal:2022,grichener:2025,morton/etal:2025,fairbairn/rafikov:2025,bhattacharyya/etal:2025} 
and dark matter backgrounds \citep{hui/etal:2017,baror/etal:2019,berezhiani/etal:2019,annulli/etal:2020,chavanis:2020,chavanis:2021,traykova/clough/etal:2021,hartman/etal:2021,Bar:2021jff,rodrigo/cardoso:2022,buehler/desjacques:2023,foote/etal:2023,tomaselli/etal:2023,traykova/vicente/etal:2023,Berezhiani:2023vlo,boudon/etal:2024,gorkavenko/etal:2025}.

Gas dynamical friction plays in key role in binary formation in gaseous environments \cite{Tagawaetal2020,Rozneretal2023,DodiciTremaine2024,Takatsyetal2025}. For example, in the active-galactic-nuclei (AGN) channel for gravitational-wave source formation, tight black-hole binaries are formed by dissipating their excess energy to the AGN disc \cite[e.g.][]{Tagawaetal2020,GonerozovPerets2023,Rownetal2023,Rowanetal2024,Whiteheadetal2024,Whiteheadetal2025}. It is dynamical friction which acts as the coupling mechanism between the binary and the energy reservoir (the disc) that enables this energy dissipation. Naturally, binary orbits, which start unbound and later lose enough energy to the gas to form a binary, are bound to be eccentric \cite[e.g.][]{GonerozovPerets2023}; therefore an understanding of dynamical friction on eccentric orbits is crucial to modelling this binary-formation mechanism. 

In this work, we consider the dynamical friction (DF) experienced by point masses on eccentric trajectories, a process relevant to many astrophysical systems.
Although curvilinear orbits have received some attention in the literature, most previous work has focused on circular orbits \cite{tremaine/weinberg:1984,kim/kim:2007,kim/etal:2008,desjacques/etal:2022,eytan/etal:2024}, with the exception of 
\cite{buehler/desjacques:2023,oneill/etal:2024}.
We concentrate on the evaluation of orbit-averaged dissipation rates, which is significantly simpler than computing the instantaneous gravitational drag force.
To this end, we develop an approach based on a frequency and multipole decomposition, valid for perturbers on generic Keplerian orbits embedded in a homogeneous, fluid-like medium 
with negligible self-gravity.
While the formalism is currently limited to linear response theory, it remains self-consistent throughout.
Our expressions for the steady-state energy and angular momentum dissipation rates eliminate the need for semi-numerical simulations of the linear density response.
They can be directly applied to study the effect of DF on the secular evolution of compact, eccentric binaries in any astrophysical environment that admits a fluid description. 

The paper is organized as follows. 
In section \ref{sec:theory}, we spell out the derivation of steady dissipation rates of orbital energy and angular momentum for a single perturber with arbitrary eccentricity, 
and compare our results to those of \cite{desjacques/etal:2022,eytan/etal:2024} in the specific case of a circular orbit. 
In section \ref{sec:sims}, we introduce the numerical simulations used to validate our theoretical predictions in the case of an ideal fluid, and discuss also the convergence 
of the simulations to a steady dissipation state. 
In section \ref{sec:BBH}, we extend our harmonic approach to the case of two perturbers on an eccentric Keplerian orbit, and apply our results to calculate the orbital evolution 
of compact, unequal-mass eccentric binaries under dynamical friction. We conclude in section \ref{sec:conclusions}. 

\section{Theory}
\label{sec:theory}

We consider an (infinite) uniform and isotropic medium characterized by a (Fourier space, retarded) Green's function $G(\omega,k)$ where $k=|\vk|$ is the wavenumber.
For example, $G(\omega,k) = \big(c_s^2 k^2 - (\omega+\mathrm{i}\varepsilon)^2\big)^{-1}$ in the case of an ideal fluid with negligible self-gravity.
This medium is perturbed by a point mass moving on a fixed eccentric orbit in the $x-y$ plane with orbital frequency $\Omega>0$, semi-major axis $a$ and eccentricity $e$. 
The point mass has a density $\rho_p(t,\vr) = m_p\, h(t)\,\delta^D(\vr-\vr_p(t))$, where the time-dependent function $h(t)$ is unity (zero) when the perturbation is turned on (off) 
and  $\vr_p(t)$ describe its periodic motion. To facilitate the analysis, we define also an orbit-averaged Mach number $\mach\equiv \Omega a/c_s$.

\subsection{Density wake}

The gravitational interaction between the moving perturber and the medium generates a density wake in the medium, which propagates outwards. 
Let $\phiw$ be the gravitational potential sourced by the density wake and $\delw$ be the overdensity of the background (the wake density). 
The gravitational potential and the wake density are related through Poisson's equation
\begin{equation}
    \nabla^2\phiw =4\pi G\delw \;.
\end{equation}
To find an expression for $\phiw$, we start from the perturbation $\delw$, which can be extracted from the fluid equations of motion using Green's theorem.

In the case of an ideal, non-gravitating fluid for instance, the linearized fluid equations yield the driven sound wave equation
\begin{align}
    \label{eq:idfluideq}
    \frac{\partial^{2}\delw}{\partial t^{2}}-c_{s}^{2}\nabla^{2}\delw=4\pi G\bar{\rho_{g}}\rho_p
\end{align}
where the right-hand side encodes the gravitational disturbance of the perturber. 
For the same point-like perturber but a general fluid-like medium, it can be shown that the linear response wake density takes the form
\begin{align}
\label{eq:deltarho}
    \delw(t,\vr)&=4\pi Gm_p\bar\rho_g\int_\omega\int_{-\infty}^{+\infty}\mathrm{d}t'h(t')\mathrm{e}^{-\mathrm{i}\omega(t-t')} \nonumber\\
    &\qquad \times\int_{\vk} G(\omega,k)\,\mathrm{e}^{\mathrm{i}\vk\cdot(\vr-\vr_p(t'))}\;,
\end{align}
where $\int_\omega \equiv \frac{1}{2\pi}\int_{-\infty}^{+\infty}\!\mathrm{d}\omega$ and $\int_{\vk} \equiv \frac{1}{(2\pi)^3}\int\!\mathrm{d}^3k$
for shorthand convenience.
Furthermore, the function $h(t)$ encodes whether the perturber is active or inactive. 
In a steady state we have $h=1$ at all times whereas, for a finite duration or step-like perturbation turned on at $t=0$, 
we have $h(t)=1$ for $t > 0$ and zero otherwise.

Let us Fourier expand the gravitational potential induced by the trailing density wake:
\begin{equation}
  \phiw(t,\vr) = \int_\omega \int_{\vk} \! \phiw(\omega,\vk) \,\mathrm{e}^{\mathrm{i}(\vk\cdot\vr -\omega t)} \;.
\end{equation}
Using Poisson's equation in Fourier space, the gravitational potential of the wake is given by:
\begin{align}
  \label{eq:phiwFT}
  \phiw(\omega,\vk) &= -(4\pi G)^2 m_p\bar\rho_g  \\
  &\qquad \times \int_{-\infty}^{+\infty}\!\mathrm{d}t'\, \frac{h(t')}{k^2} G(\omega,k)\, \mathrm{e}^{-\mathrm{i}\vk\cdot\vr_p(t')+\mathrm{i}\omega t'} \nonumber \;.
\end{align}
The force acting on the perturber can be easily read as:
\begin{align}
\label{eq:wakeforce}
  {\bf F}_p(t) &= (4\pi Gm_p)^2 \bar\rho_g \int_\omega\int_{-\infty}^{+\infty}\! \mathrm{d}t'\,h(t')\,\mathrm{e}^{-\mathrm{i}\omega (t-t')} \nonumber \\
  &\qquad \times \int_{\vk}\!\frac{\mathrm{i}\vk}{k^2}\, G(\omega,k)\,
  \mathrm{e}^{\mathrm{i}\vk\cdot(\vr_p(t)-\vr_p(t'))} \\
  & = - m_p\grad_{\vr}\phiw(t,\vr)\bigg\lvert_{\vr=\vr_p(t)} \nonumber \;.
\end{align}
It is proportional to $(Gm_p)^2$ because the gravitational field of the perturber does work on the fluid (and launches the density wake), and the 
density disturbances exert a gravitational pull on the perturber.

Finally, note that all these results are valid in linear response theory, which we will assume throughout this paper.

\subsection{Orbital energy and angular momentum dissipation}

We calculate the work done by the wake on the perturber during an orbital revolution (defined as the time interval between two pericentre passages)
starting from
\begin{equation}
  \mathrm{d}\phiw = \partial_i \phiw \mathrm{d}x^i + \dphiw \mathrm{d}t \;,
\end{equation}
where $\dphiw = \frac{\partial}{\partial t}\phiw$, 
and integrate $d\phiw$ on a path $\gamma$ going from $(t_1,\vr_p(t_1))=(t_1,\vr_1)$ to $(t_2,\vr_p(t_2))=(t_2,\vr_2)$. 
Here, $\vr_p(t)$ describes the quasi-periodic motion of the perturber, such that $\vr_p(t+T)\approx \vr_p(t)$ where $T=2\pi/\Omega$ is the orbital period.
Therefore, upon setting $t_2 = t_1+T$, we get
\begin{align}
  \int_{\partial \gamma}\! \phiw &= \phiw(t_2,\vr_2) - \phiw(t_1,\vr_1) \\
  &\approx \phiw(t_1+T,\vr_1) - \phiw(t_1,\vr_1) \nonumber 
\end{align}
so long as the DF force is a small perturbation to the perturber's trajectory. 
This is equivalent to the Born approximation in gravitational lensing (the effect of the perturbation is computed along the unperturbed trajectory).
In steady state, $\phiw(t_1+T,\vr_1) = \phiw(t_1,\vr_1)$  and the integral vanishes.
In this regime, Stokes' theorem 
\begin{equation}
  \int_\gamma\! \mathrm{d}\phi = \int_{\partial \gamma}\! \phi 
\end{equation}
thus implies
\begin{equation}
  \int_\gamma\! \mathrm{d}\phiw = \int_\gamma\! \mathrm{d}x^i\, \partial_i\phiw + \int_\gamma\! \mathrm{d}t\, \dphiw = 0
\end{equation}
in the Born approximation.
Since $-\partial_i\phiw$ is the force per unit mass acting on the perturber, the orbit-averaged rate of energy dissipated due to dynamical friction is 
\begin{equation}
  \label{eq:Edot}
  \langle\dot E\rangle = \frac{m_p\Omega}{2\pi} \int_\gamma\! \mathrm{d}t\, \frac{\partial}{\partial t} \phiw \;.
\end{equation}
This is equal to the work done by the sound wake on the perturber (dynamical friction). 
Note that $\dphiw\propto \dot{\rho}_P$, i.e. the time-dependence of $\phiw$ arises only through the motion of $m_p$.

Likewise, the torque in the direction tranverse to the orbital plane is
\begin{equation}
    N_z = m_p \Big(-\vr_p\times \grad_{\vr_p}\phiw\Big)_z = \frac{\mathrm{d}L_z}{\mathrm{d}t} \;.
\end{equation}
Therefore, the orbit-averaged dissipation rate of the $z$-component of the angular momentum is
\begin{equation}
    \langle\dot{L}_z\rangle = \frac{m_p\Omega}{2\pi} \int_\gamma\! \mathrm{d}t\, \hat L_z\, \phiw\;,
\end{equation}
where $\hat L_z \equiv \big(-\vr_p\times \grad_{\vr_p}\big)_z$.
In general, we would also need
\begin{equation}
    \left\langle\frac{\mathrm{d}}{\mathrm{d}t} L^2\right\rangle = \frac{m_p\Omega}{2\pi} \int_\gamma\! \mathrm{d}t\, \hat L^2\, \phiw
\end{equation}
with $\hat L^2\equiv \big(\vr_p\times \grad_{\vr_p}\big)^2$.
However, since in our setting ${\bf L}$ is oriented along the $z$-axis, $\langle \dot{L}_z\rangle$ suffices to characterize the dissipation of angular momentum.

\subsection{Steady-state expressions}

\label{sub:steady}

Steady-state expressions are characterized by 3 integers $(j,l,m)$, which encode the energy and angular momentum of the sound waves.

On setting $h(t)\equiv 1$ and substituting Eq.~(\ref{eq:phiwFT}) into Eq.~(\ref{eq:Edot}), we get
\begin{align}
    \label{eq:Me}
    \langle\dot E\rangle &= -(4\pi Gm_p)^2 \bar\rho_g\frac{\Omega}{2\pi} \int_0^{\frac{2\pi}{\Omega}}\!\mathrm{d}t \int_{-\infty}^{+\infty}\! \mathrm{d}t' \\
    &\qquad\times\int_\omega\int_{\vk} \,\frac{(-\mathrm{i}\omega)}{k^2}\, G(\omega,k)\mathrm{e}^{\mathrm{i}\vk\cdot(\vr_p(t)-\vr_p(t'))-\mathrm{i}\omega (t-t')} \nonumber \;.
\end{align}
Exploiting the periodicity of the unperturbed trajectory $\vr_p(t)$ and decomposing the result into harmonic modes (see Appendix \S\ref{app:math} for details), 
the orbit-averaged energy dissipation rate can also be expressed as
\begin{align}
\label{eq:Edotkintegral}
 \langle\dot E\rangle &= -32\pi(Gm_p)^2 \bar\rho_g\Omega  \\
 &\qquad \times \sum_{j\ne 0,l,m} j\, \Im\left\lbrace\int_0^\infty\!\mathrm{d}k\, G(j\Omega,k)\,\big\lvert I_{j\ell m}(k,e)\big\lvert^2\right\rbrace 
 \nonumber \;,
\end{align}
where $\Im(z)$ is the imaginary part of $z$ and the auxiliary function $I_{j\ell m}(k,e)$ is given by
\begin{equation}
    \label{eq:Ijlmk}
     I_{j\ell m}(k,e) = \frac{1}{2\pi}\int_0^{2\pi}\!\mathrm{d}M\,j_\ell\big(k r_p(M)\big)\,Y_\ell^m\big(\rvh_p(M)\big)\,\mathrm{e}^{-\mathrm{i}\,jM}
\end{equation}
For a bound Keplerian orbit, the mean and eccentric anomalies, $M$ and $\xi$, are related through $M= \xi-e\sin\xi$. 

The calculation of $\langle \dot L_z\rangle$ proceeds analogously. Using
\begin{equation}
    \hat L_z\, Y_\ell^m(\rvh_p) = -i m\, Y_\ell^m(\rvh_p) \;,
\end{equation}
we obtain
\begin{align}
\label{eq:Ldotkintegral}
 \langle \dot{L}_z\rangle &= -32\pi(Gm_p)^2 \bar\rho_g \\
 &\qquad \times \sum_{j\ne 0,l,m} m\,\Im\left\lbrace\int_0^\infty\!\mathrm{d}k\,G(j\Omega,k)\,\big\lvert I_{j\ell m}(k,e)\big\lvert^2\right\rbrace \nonumber \;.
\end{align}
This expression is identical to $\langle \dot E\rangle$ except for $j\Omega$ replaced by $m$. 
Note that, in equations~(\ref{eq:Edotkintegral}) and (\ref{eq:Ldotkintegral}), the sum over $j$ excludes $j\ne 0$ (which does not correspond to a propagating
sound wave) while there is no such restriction on $(\ell,m)$.

It is convenient to express the orbit-averaged energy and angular momentum dissipation in terms of the friction coefficients $I_E$ and $I_L$, which we define according to
\begin{align}
    \label{eq:coeffriction}
    \langle\dot E\rangle &= -  4\pi \frac{\left(Gm_p\right)^2}{\Omega a} \bar\rho_g\, I_{E}(\mach,e) \\
    \langle\dot{L}_z\rangle &= -  4\pi \frac{\left(Gm_p\right)^2}{\Omega^2a} \bar\rho_g\, I_{L}(\mach,e) \nonumber \;.
\end{align}
Note that $I_E\ne I_L$ when $e>0$.

The steady-state solutions for $\langle\dot E\rangle$ and $\langle\dot{L}_z\rangle$ depend on the orbital geometry encoded in the
quantity $I_{j\ell m}(k,e)$ and the physical characteristics of the density wake encoded in the Green's function $G(\omega,k)$.

\subsection{Computing the geometrical form factor $I_{j\ell m}$}
\label{sec:Ijlm}

For convenience, we write 
\begin{equation}
    I_{j\ell m}(k,e) \equiv Y_\ell^m\!\Big(\frac{\pi}{2},0\Big)\, \gjlm (ka, e)
\end{equation}
where the functions
\begin{equation}
\label{eq:gjlm}
    \gjlm(x, e) \equiv \frac{1}{2\pi}\int_{-\pi}^{\pi}\!\mathrm{d}M\,j_\ell\big(x (r_p(M)/a)\big)\,\mathrm{e}^{\mathrm{i}(m\varphi-jM)}
\end{equation}
satisfy $\gjlm(x,e)=g^{(-j)}_{\ell, -m}(x,e) =(-1)^\ell\, \gjlm(-x,e)$. 
We consider two approximations to $\gjlm(x,e)$, which can be validated through a direct numerical integration of Eq.~(\ref{eq:gjlm}).

For small values of $x=ka$, we insert in Eq.~(\ref{eq:gjlm}) the series expansion of the spherical Bessel function $j_\ell(z)$ around $z=0$.
This leads to integrals of the form 
\begin{equation}
\label{eq:Intnmj}
    \int_{0}^{2\pi}\mathrm{d}M\left(\frac{r_{p}}{a}\right)^{n}\mathrm{e}^{\mathrm{i}m\varphi}\mathrm{e}^{-\mathrm{i}jM}=2\pi X_j^{n,m}(e) \;,
\end{equation}
where the symbols $X_s^{n,m}(e)$ denote the Hansen coefficients \cite{hughes:1981,murray/dermott:1999,mardling:2013}, whose basic properties are 
summarized in Appendix \S\ref{app:hansen}. The Hansen coefficients enable us to recast $\gjlm(x,e)$ into the series expansion
\begin{align}
    \label{eq:gjlmhansen}
    g^{(j)}_{\ell m}(x, e) 
    &= x^\ell \sum_{q=0}^\infty \frac{(-1)^q}{q! (2q+2\ell+1)!!}\left(\frac{x^2}{2}\right)^qX_j^{2q+\ell,m}(e)  
    \nonumber \\
    &= j_\ell(x) \delta_{j m} + \mathcal{O}(e) \;.
\end{align}
For integer values of $\ell$, this series defines an entire function of $x$. 
We truncate it at some finite cutoff $q_{\text{max}}$ for practical evaluations.

For large values of $x=ka$, the function $\gjlm(x,e)$ can be evaluated using the asymptotic expansion for spherical Bessel functions together 
with the stationary phase approximation for the computation of the integral over the mean anomaly in Eq.~(\ref{eq:gjlm}).
The detailed calculation reported in Appendix \S\ref{app:phase} yields 
\begin{align}
    \label{eq:gjlmspa}
    g_{\ell m}^{(j)}(x,e)&=\frac{1}{\sqrt{2\pi e}x^{3/2}}\bigg[\sin\!\Big(x(1-e)-\frac{l}{2}\pi+\frac{\pi}{4}\Big) \nonumber \\
    &\qquad +(-1)^{j+m}\sin\!\Big(x(1+e)-\frac{l}{2}\pi-\frac{\pi}{4}\Big)\bigg ] \nonumber \\
    & \quad + \mathcal{O}\big(x^{-5/2}\big) \;.
\end{align}
We retain only the leading term of this asymptotic expansion, 
which corresponds to a superposition of two decaying sinusoidal functions with wavenumber $k\propto (1\pm e)$. 

\begin{figure}
    \centering
    \includegraphics[width=0.45\textwidth]{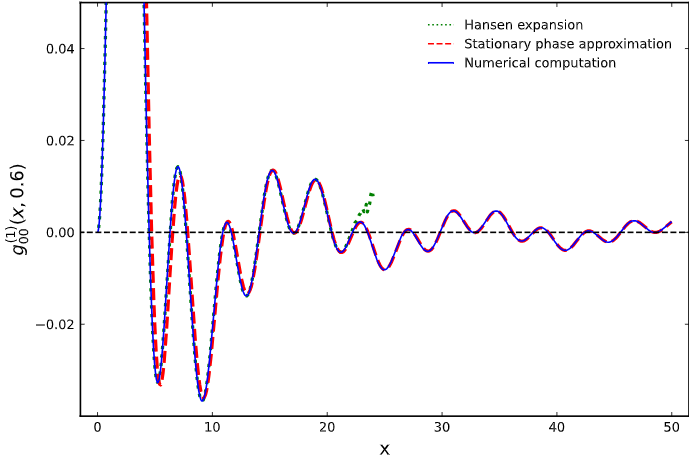}
    \includegraphics[width=0.45\textwidth]{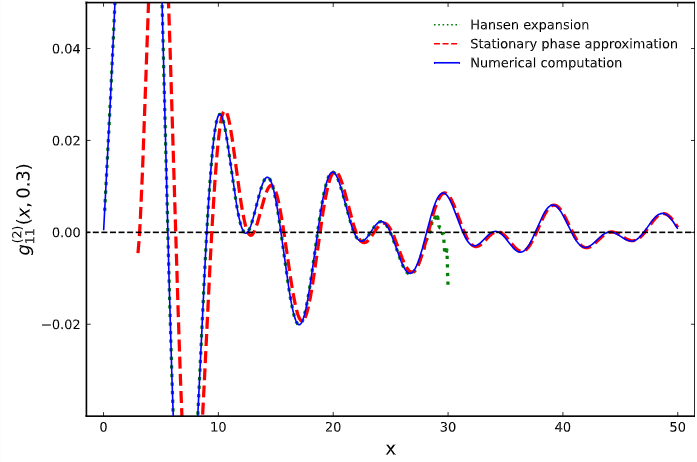}
    \caption{Comparison between the geometrical form factor $\gjlm(x,e)$ calculated using a direct numerical integration (blue solid curve), 
    the Hansen expansion (\ref{eq:gjlmhansen}) (green dotted curve) and the stationary phase approximation (\ref{eq:gjlmspa}) (red dashed curve). 
    Results are shown for $(j,\ell,m)=(1,0,0)$ and $e=0.6$ in the top panel, whereas $(j,\ell,m)=(2,1,1)$ and $e=0.3$ are assumed in the bottom panel.}
    \label{fig:gjlm}
\end{figure}

In Fig.~\ref{fig:gjlm}, these approximations are compared to a direct numerical integration of Eq.~(\ref{eq:gjlm}). 
A fiducial value of $\sigma_\text{max}=70$ is assumed.
The stationary phase approximation works best for large values of $x$, while the Hansen approximation (\ref{eq:gjlmhansen}) is accurate for small arguments.
At fixed $\sigma_\text{max}$, the latter fails to converge at smaller values of $x$ as the eccentricity $e$ increases.

\subsection{The case of an ideal fluid}

For the sake of concreteness, let us specialize our results to an ideal fluid of density $\bar\rho_g$ and (adiabatic) speed of sound $c_s$.
Unlike the collisionless case \cite{chandrasekhar:1943}, where energy and angular momentum are transferred through two-body gravitational encounters, 
in the fluid case they are transferred via the excitation of collective modes, such as sound waves.

\subsubsection{Friction coefficients}

Consider first the circular case, for which $r_p=a$ and the unit separation vector is $\rvh_p(t)=\cos(\Omega t)\,\xvh +\sin(\Omega t)\,\yvh$. This yields
\begin{align}
    \label{eq:Ijlmkcircsingle}
    I_{j\ell m}(k,e=0) &= j_\ell(ka)\,Y_\ell^m\big(\frac{\pi}{2},0\big) \,
    \frac{1}{2\pi}\int_0^{2\pi}\!\mathrm{d}M\,\mathrm{e}^{\mathrm{i}(m-j)M} \nonumber \\
    &= j_\ell(ka)\,Y_\ell^m\!\Big(\frac{\pi}{2},0\Big)\,\delta_{mj} \;.
\end{align}
Substituting the Green's function of a homogeneous and isotropic ideal fluid into the $k$-integral of Eqs.~\eqref{eq:Edotkintegral},\eqref{eq:Ldotkintegral}, we are left with a $k$-integral of the form
\begin{align}
    \label{kintcircular}
    \int_0^\infty\!\mathrm{d}k\,&\frac{j_\ell(ka)j_\ell(ka)}{c_s^2k^2 - (j\Omega+\mathrm{i}\varepsilon)^2} \\
    &= \frac{a}{c_s^2} \frac{\mathrm{i}\pi}{2}\frac{\left(j_{\ell+1}(j\mach)h_\ell^{(1)}\!(j\mach) + j_\ell(j\mach) h_{\ell-1}^{(1)}\!(j\mach)\right)}{2\ell+1}
    \nonumber \;.
\end{align}
This result was obtained from the Residue theorem after decomposing $j_\ell(x)$ into an incoming and an outgoing wave, 
i.e. $j_\ell(x) = \frac{1}{2}(h_\ell^{(1)}\!(x)+h_\ell^{(2)}\!(x))$, 
and taking into account the resonant poles $ka=\pm(\mach+\mathrm{i}\varepsilon)$ and a pole at the origin. 
Importantly however, the latter does not contribute to $\langle\dot E\rangle$ (because it yields a purely real contribution to the $k$-integral). 
Therefore, the friction coefficients satisfy
\begin{align}
    \label{eq:Icirc}
    I_E(\mach,0) &= I_L(\mach,0) \\
    &= \mach \sum_{\ell=1}^{\lmax}\sum_{m\ne 0} c_{\ell m} \Big(j_\ell(m\mach)\Big)^2
    \nonumber \;,
\end{align}
where the symbols
\begin{equation}
    \label{eq:clm}
    c_{\ell m}\equiv (2\ell+1)\frac{(\ell-m)!}{(\ell+m)!}\big\lvert P_\ell^m(0)\big\lvert^2 
\end{equation}
are introduced for shorthand convenience. Note that $c_{\ell m}$ is invariant under reflection $m\to -m$.
Eq.~(\ref{eq:Icirc}) precisely matches the tangential friction coefficient obtained for the circular case \cite{desjacques/etal:2022}. Note also that the rates satisfy $\langle\dot E\rangle = \Omega \langle\dot{L}_z\rangle$, which follows from $E=-\frac{1}{2}\Omega L_z$ with $L_z=L=m_pa^2 \Omega \equiv m_p (Gm_t)^{1/3}\Omega^{-1/3}$ (Here, $m_t\gg m_p$ is the mass of a fictitious object located at the origin, which would give rise to the correct Kepler's laws for the single eccentric object we are considering).

In the limit $\mach\to 0$, we have $\frac{1}{x}j_1(x)\to \frac{1}{3}$ which implies $I_E(\mach)\sim \frac{1}{3}\mach^3$. In fact, a small-$\mach$ development shows that, for $\lmax\to\infty$, 
\begin{align}
    I_E(\mach) &= \frac{\mach^3}{3}+ \frac{\mach^5}{5} + \frac{\mach^7}{7} + \dots \\&={\rm arctanh}(\mach) -\mach = \frac{1}{2}\ln\left(\frac{1+\mach}{1-\mach}\right) - \mach \;\nonumber,
\end{align}
which converges for $|\mach|<1$ (this was also noted in \cite{Berezhiani:2023vlo}). 
Hence, in the subsonic regime, the steady-state circular solution precisely agrees with the finite time linear expression \cite{ostiker:1999}, with the caveat that the latter is singular at $\mach=1$ while Eq.~(\ref{eq:Icirc}) converges for any $\mach\geq 0$ so long as the multipole expansion is truncated at some finite $\lmax$.

Consider now the eccentric case. 
The function $\gjlm(z,e)$, like $j_\ell(z)$, is holomorphic in the whole complex plane for integer $\ell$ and admits a singularity at infinity along the imaginary axis.
A decomposition of $\gjlm(z, e)$ into the eccentric analog of outgoing and incoming spherical waves 
yields a pole at $k=0$ in the $k$-integrals of Eqs.~(\ref{eq:Edotkintegral}) and (\ref{eq:Ldotkintegral}). However, this pole result in a purely real contribution to the $k$-integral and, therefore, a vanishing contribution to $\langle\dot E\rangle$ and $\langle\dot{L}_z\rangle$. For an eccentric perturber, the friction coefficients are thus determined by the resonant poles $c_sk=\pm j \Omega$ solely, in complete analogy with the circular case. They can be recast into
\begin{align}
    \label{eq:Ieccentric}
    I_E(\mach,e) &=\mach \sum_{j\ne 0}\sum_{\ell,m} c_{\ell m} \Big(g^{(j)}_{\ell m}(j\mach,e)\Big)^2 \;, \\
    I_L(\mach,e) &=\mach \sum_{j\ne 0}\sum_{\ell,m} \frac{m}{j} c_{\ell m} \Big(g^{(j)}_{\ell m}(j\mach,e)\Big)^2\nonumber \;.
\end{align}
We have $I_E\geq 0$ always since $c_{\ell m}\geq 0$.
To see that $I_L\geq 0$ (when $L_z>0$ as assumed here), 
note that the friction coefficient associated with $\langle \frac{\mathrm{d}}{\mathrm{d}t}L^2\rangle$ is identical to $I_L$ except for the replacement $m\to l(l+1)$.
Finally, the condition $j\ne 0$ ensures that, when $e>0$, both $I_E(\mach,e)$ and $I_L(\mach,e)$ are proportional to $\mach^3$ in the limit $\mach\ll 1$, 
as in the circular case.

\subsubsection{Wake density}

Our approach can also be used to calculate explicitly the density wake launched in the ideal fluid by the eccentric perturber.
Appendix \S\ref{app:wake} shows that, in the steady-state regime, 
the wake density perturbation is given by the following superposition of outgoing sound waves, 
\begin{align}
    \label{eq:delrhotint}
    &\delw(t,\vr) =2\mathrm{i}\bar\rho_g \mach \frac{r_B}{a}\sum_{j\ne0}\sum_{lm}j\,Y_{\ell}^{m}\left(\rvh\right)Y_{\ell}^{m*}\!\left(\frac{\pi}{2},0\right)\\&\times \mathrm{e}^{-\mathrm{i}j\Omega t}\int_{0}^{2\pi}\mathrm{d}M'\mathrm{e}^{\mathrm{i}(jM'-m\varphi')}j_\ell\!\left(j\mathcal{M}\frac{r_{<}}{a}\right)h_{\ell}^{(1)}\!\left(j\mathcal{M}\frac{r_{>}}{a}\right) \nonumber\\
    &+\bar{\rho}_{g}r_{B}\sum_{lm}\int_{0}^{2\pi}\mathrm{d}M'\frac{1}{2\ell+1}\frac{r_{<}^{\ell}}{r_{>}^{\ell+1}}Y_{\ell}^{m}(\rvh)Y_{\ell}^{m*}(\rvh_{p}')\nonumber
\end{align}
where $r_B=2 Gm_p/c_s^2$ is the Bondi radius of the perturber, $\varphi'=\varphi(M')$ is the azimuthal angle, $r_<=\min(r,r_p')$ and $r_>=\max(r,r_p')$.
The second term in the right-hand side of Eq.~(\ref{eq:delrhotint}) is the contribution from $j=0$.

For $e=0$, the radial coordinate satisfies $r_p=a$ independently of time and the $M'$-integral returns a Kronecker delta $\delta_{jm}$. 
As a result, we recover the circular steady-state expression \cite{eytan/etal:2024} 
\begin{align}
    \delw^{\text{circ}}(t,\vr)&=2\pi \mathrm{i}\bar\rho_g \mach \frac{r_B}{a}\sum_{l,m\ne0}m\, j_\ell\!\left(m\mathcal{M}\frac{r_<}{a}\right)\\
    &\qquad\times h_{\ell}^{(1)}\!\!\left(m\mathcal{M}\,\frac{r_>}{a}\right)Y_{\ell}^{m}(\rvh)\, Y_{\ell}^{m*}(\rvh_p(t))\nonumber\;\\
    &\quad +\frac{1}{2}\bar{\rho}_{g}r_{B}\sum_{l}\frac{r_{<}^{\ell}}{r_{>}^{\ell+1}}|P_{\ell}^{0}(0)|^{2}\nonumber \;,
\end{align}
with $r_<=\min(r,a)$ and $r_>=\max(r,a)$.
For eccentric orbits, the density perturbation at a radial coordinate $r>\rapo=a(1+e)$ larger than the apocentre distance can be computed using the fact 
that $r_<=r_p$ for all $t'$. 
Expressing the $M'$-integral in terms of $\gjlm(j\mach,e)$ yields a sound wake density
\begin{align}
    \label{eq:delrhoR}
    \delw^{r>\rapo}&=2\pi \mathrm{i}\bar\rho_g \mach \frac{r_B}{a}
    \sum_{j}\sum_{lm}j\,Y_{\ell}^{m}(\rvh)\,Y_{\ell}^{m*}\!\left(\frac{\pi}{2},0\right) \nonumber \\
    &\qquad \times \mathrm{e}^{-\mathrm{i}j\Omega t}{\gjlm}^*\!(j\mach,e)\, h_{\ell}^{(1)}\!\!\left(j\mathcal{M}\frac{r_{>}}{a}\right)\\
    &\quad +2\pi\bar{\rho}_{g}r_{B}\sum_{lm}\frac{X_{0}^{\ell,m}}{2\ell+1}\frac{a^{\ell}}{r^{\ell+1}}Y_{\ell}^{m}(\rvh)Y_{\ell}^{m*}\!\left(\frac{\pi}{2},0\right)
    \nonumber
\end{align}
outside the orbit falling off as $r^{-1}$~\footnote{This can be seen by considering the monopole ($\ell=0$) contribution and substituting the spherical Hankel and 
Bessel functions with their asymptotic expansions.}, as in the circular case.
In Section \S\ref{sub:acousticwake} below, we confirm that these expressions match very well simulations.

\section{Validation with simulations}
\label{sec:sims}

We test the validity of our analytical expressions with numerical simulations of the linear response of an ideal, homogeneous and isotropic fluid.

\subsection{Simulation setup}

\label{sub:setup}

To simulate the acoustic wake induced by a point-like perturber on an eccentric Keplerian orbit, we perform three-dimensional, finite-difference 
time-domain (FDTD) simulations on a cubic grid. These simulations solve the linearized continuity and momentum conservation equations
\begin{align}
    \dot\delw+\bar\rho_g\grad^2\psi &=0 \\
    \dot\psi+c_s^2\frac{\delw}{\bar\rho_g}&=-\phi_p \nonumber \;,
\end{align}
where $\phi_p$ is the gravitational potential of a point mass, 
for the velocity potential $\psi$ and the fluid density perturbation $\delw$, such that $\velw=\grad\psi$ is the velocity field.

The simulations grid includes $N\times N \times (N/2)$ cells owing to the symmetry of the perturbations relative to the orbital ($x-y$) plane.
By construction, these simulations implement a finite time perturbation initially located at pericentre and active for $t>0$. 
To ensure the validity of the linear response, the spatial curbical grid is chosen such that its spatial resolution 
is coarser than the Bondi radius, albeit significantly finer than the characteristic orbital scale $a$.
The perturber is assigned a gaussian density profile of width $\sigma=0.1a$, which is resolved by the grid cell spacing $\Delta x$, ranging from $0.01a$ to $0.03a$. 
The simulation time is such that the perturber completes two orbital revolutions.
We restrict ourselves to Mach numbers $0\leq\mach\leq 2.5$ to ensure that a steady state of energy and angular momentum dissipation is reached by the end of the simulations
(cf. Section \S\ref{sub:measurements}).
The grid length $L$ is chosen to fully enclose the density perturbation at all times, ensuring that boundary effects can be neglected.
In practice, the grid length is set to $2\times (c_s T+a)\times 1.05$ since the acoustic wake starts its propagation from the focal point  
$\vr=(a(1-e),0,0)$ for a period of $2T$.
The simulation time steps are constant intervals of eccentric anomaly $\Delta\xi$.
The choice $\Delta\xi=\frac{1}{2}\Omega \Delta x/(\sqrt3\max(c_s,\vpmax)(1+e))$, where $\vpmax=\Omega a \sqrt{(1+e)/(1-e)}$ 
is the maximal velocity of the perturber,
guarantees that the CFL (Courant-Friedrichs-Lewy) condition is satisfied for both the perturber's velocity and the medium sound velocity. 
In this regime, refining the resolution does not affect the results noticeably.

Our analytic framework involves multipole expansions, which are truncated in practice at a maximum angular multipole $\lmax$.
In real astrophysical systems, $\lmax$ is determined by physical scales such the size of the perturber's Bondi radius etc.
In simulations, $\lmax$ is set by the finite angular resolution. 
For a perturber located at radial distance $R$, the truncation of the multipole expansion implies that the smallest resolvable azimuthal wavelength is 
$\lambda_\text{min} = \frac{2\pi R}{\lmax}$. 
This is analogous to a spatial resolution \( \Delta x_{\text{eff}} = \frac{\pi R}{\ell_{\text{max}}} \) in the tangential direction. 
To ensure a consistent comparison between the simulated and theoretical rates of dissipation, we apply a time-dependent low-pass filter in Fourier space to the 
simulated gravitational potential, retaining only modes with wavenumber \( |\mathbf{k}| < \frac{\pi}{\Delta x_{\text{eff}}} \). 
This procedure -— which depends on the distance between the perturber and the origin -— effectively imposes the same angular resolution limit as the multipole cutoff in the analytic approach.
By contrast, the simulated radial force acting on the perturber is calculated without a low-pass filter since the radial component in the analytical formulation 
has infinite resolution (the $k$-integral in equations~(\ref{eq:Edotkintegral}) and (\ref{eq:Ldotkintegral}) runs from 0 to $\infty$). 

\begin{figure*}
    \centering
    \includegraphics[width=\textwidth]{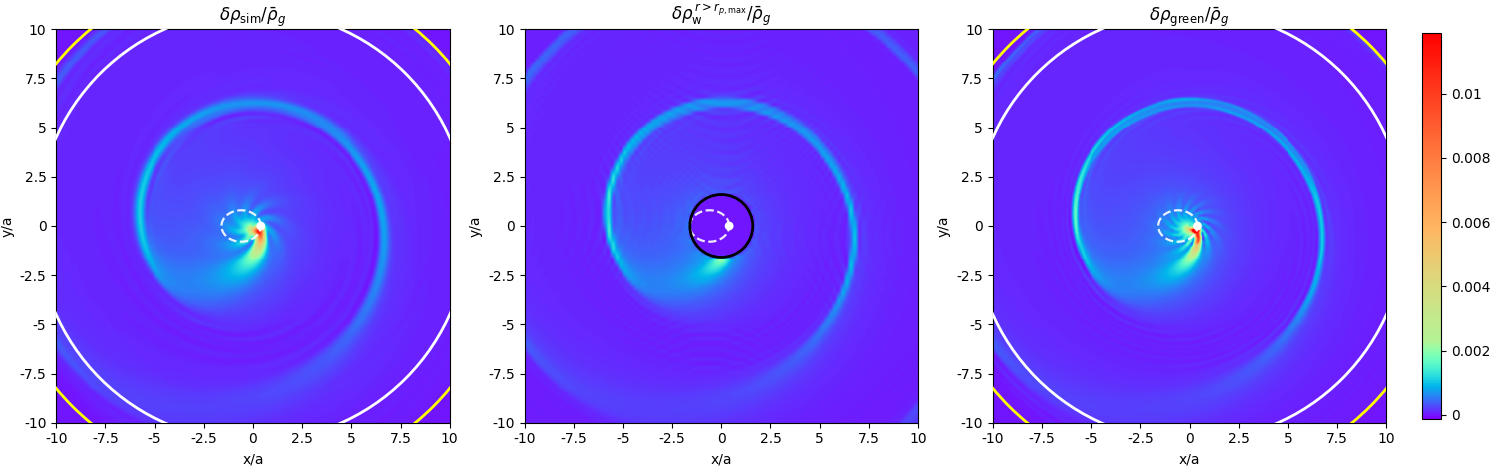}
    \caption{The acoustic wake overdensity $\delta\rho/\bar\rho_g$ in the orbital plane of a perturber with $e=0.6$ and $\mach=1$.
    {\it Left panel}: overdensity extracted from the FDTD simulation after two orbital revolutions, decomposed via a multipole expansion up to 13th order. 
    {\it Middle panel}: steady state overdensity, computed by Eq.~(\ref{eq:delrhoR}) with the cutoffs $\ell_{\text{max}}=13$ and $j_{\text{max}}=20$. 
    {\it Right panel}: overdensity after two complete cycles computed with the roots methods decomposed with multipole expansion up to the 13th order. 
    In all panels, the white, dashed ellipse and the white dot denote the Keplerian path of the perturber and its current location, whereas the outer (yellow) 
    and inner (white) circles mark the boundaries of the steady-state and information sphere, respectively.}
    \label{fig:delrho}
\end{figure*}    

Our analytic expressions Eq.~(\ref{eq:Ieccentric}) involve also a summation over $j$, which is restricted to the range $-\jmax\leq j\leq \jmax$ for
practical evaluations. The value of $\jmax$ is set according to the maximum frequency $\omega_\text{max}=\jmax \Omega$ resolved by the simulations, 
for which we can get a rough upper bound using $\omega_\text{max} \sim \Delta t^{-1}$ with $\Delta t = (1-e\cos\xi) \Omega^{-1}\Delta\xi$. 
Our choice of $\Delta\xi$ yields a minimum time interval
\begin{equation}
    \Delta t \gtrsim \frac{\Delta x}{(1+e)^2\, \Omega a\, {\rm max}\!\left(\mach^{-1},\sqrt{\frac{1+e}{1-e}}\right)}
\end{equation}
and, therefore,
\begin{equation}
    \label{eq:jmax}
    \jmax \lesssim \bigg(\frac{a}{\Delta x}\bigg) \big(1+e\big)^2{\rm max}\!\left(\mach^{-1},\sqrt{\frac{1+e}{1-e}}\right)
\end{equation}
Our simulation setting with a resolution $a/\Delta x \sim 10 - 10^2$ thus implies, in principle, $\jmax\sim \mathcal{O}(10 - 10^2)$ with a noticeable dependence 
on the Mach number (mainly in the subsonic regime) and the eccentricity.

\subsection{Acoustic wake}

\label{sub:acousticwake}

The acoustic wake never reaches a steady state since the wavefronts emerging from the perturber, which determine the "information sphere", have a finite radius bounded by $c_s t$ at time $t$. 
Nevertheless, we can still ask how the density wake behaves inside this sphere. 
The wake velocity field $\velw$ can be calculated from the linearized momentum conservation equation, 
\begin{equation}
    \frac{\partial\velw}{\partial t}=-c_{s}^{2}\grad\frac{\delw}{\bar\rho_g}-\frac{G m_p}{r_p^2(t)}\rvh_p(t) \;.
\end{equation}
For a finite time perturbation turned on at $t=0$, a steady-state density wake can be established only at times $t>a(1+e)/c_s$, within a sphere 
of radius $r=c_s t - a(1+e)$ (see Appendix \S\ref{app:wake}).

The steady-state acoustic wakes extracted from the simulations and computed with the direct method (Eq. (\ref{eq:delrhoOstriker}) are compared in Fig.~\ref{fig:delrho} 
for an eccentric perturber with $e=0.6$. Results are shown after the perturber has completed two orbital cycles. 
The simulated wake appears somewhat smoother due to numeric artifacts arising from the grid resolution of the FDTD simulations.
On applying a Gaussian filter of width $\sigma=0.05 a$ to mitigate the resolution dependence, we obtain a relative error between $\delrho_{\text{sim}}$ and $\delw^{r>\rapo}$ 
of less than $2\%$ throughout most of the steady-state region, except in small areas where $\delrho$ approaches zero.
The error between $\delrho_{\text{green}}$ and $\delw^{r>\rapo}$ is much smaller, as expected, since the direct method is spatially accurate, 
unlike the FDTD simulation which is limited by grid resolution. 

Energy is continuously injected into the wake by the perturber, which remains locked on a fixed eccentric trajectory.
The average energy flux through a sphere of radius $r$, which is second order in the perturbations, is given by
\begin{equation}
    \left\langle F\right\rangle = \int_{S}\delw\, \velw \cdot  \mathrm{d}{\bf S} \;,
\end{equation}
where $\mathrm{d}{\bf S}$ is a infinitesimal surface element. 
Far from the origin, in the limit $r\rightarrow\infty$ and $t\to\infty$ such that the wake has reached a steady state up to the radius $r$, we find 
$\langle\dot E\rangle = \langle F\rangle$.
This implies that the energy required to maintain the mass on its fixed orbit is entirely transferred to the wake. 
It is worth noting that, in this long-time limit, the total energy carried by the sound wake formally diverges because energy has been continuously 
injected for an infinite amount of time. 

\begin{figure}
    \centering
    \includegraphics[width=0.45\textwidth]{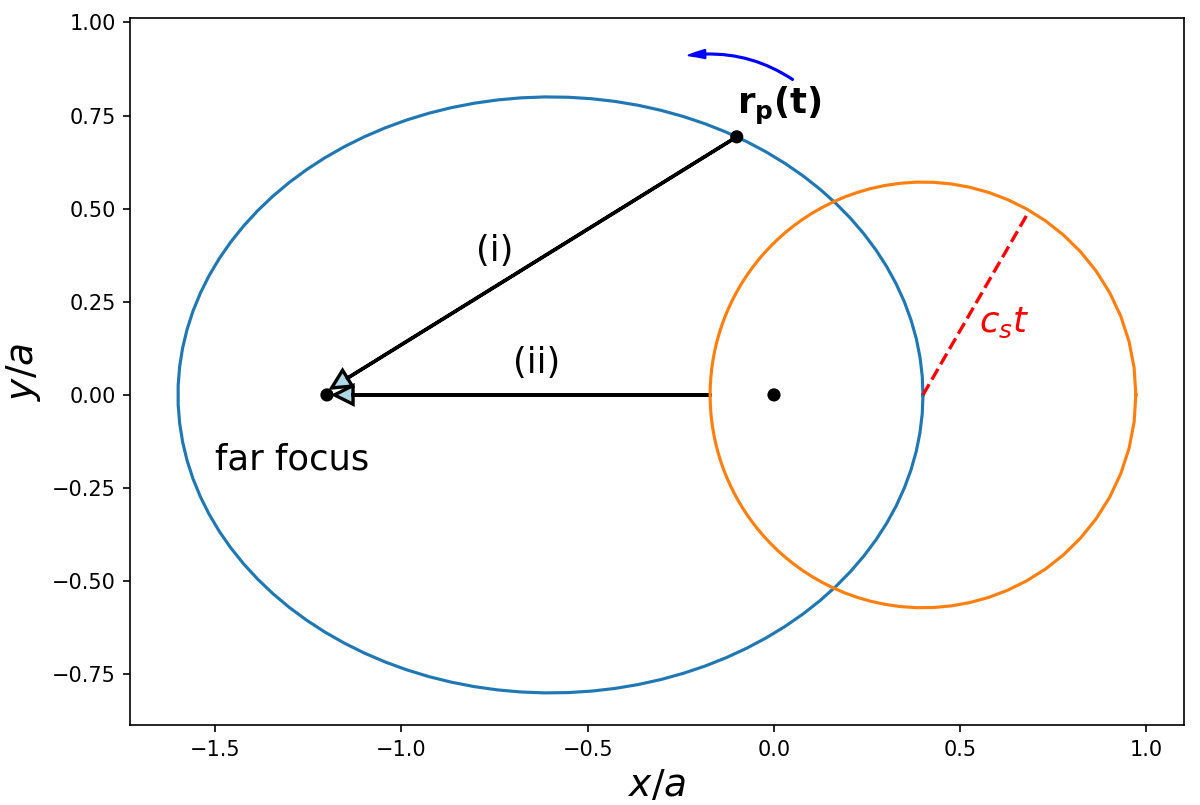}     
    \includegraphics[width=0.45\textwidth]{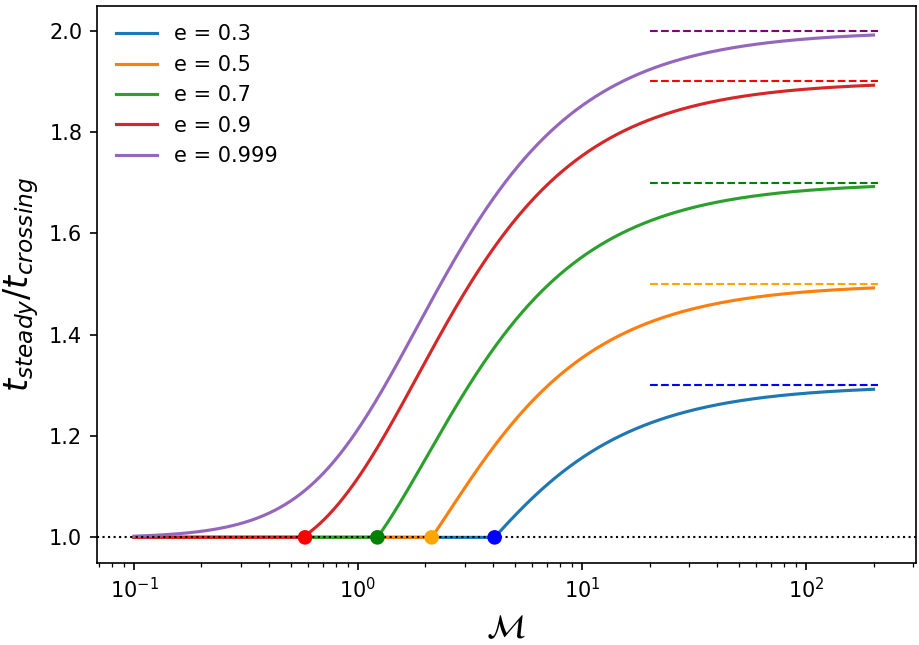}   
    \caption{{\it Upper panel}: Eccentric path of a supersonic perturber with eccentricity $e=0.6$ (blue ellipse) along with the initial information sphere coming from $a(1-e)\hat x$ 
    (orange circle). The close focus point is at the origin, while the far focus is at $-a(1+e)\hat x$. Distances $(i)$ and $(ii)$ such that $D=(i)-(ii)$ (see main text) are shown 
    with the black arrows.
    {\it Lower panel}: Characteristic timescale $t_\text{steady}$ to reach a steady dissipation state (of energy and angular momentum) 
    when the perturber is turned on at $t=0$. 
    The steady-state time $t_\text{steady}$ is computed from Eq.~(\ref{eq:tsteady}).
    It is plotted in units of the sound crossing time $t_\text{crossing}=2a/c_s$. 
    The dashed horizontal lines indicate the asymptotic behavior of the steady state time at high Mach number, where $t_{\text{steady}}\to (1+e)t_{\text{crossing}}$. 
    The circles mark the breaking points occuring at the critical Mach number $\mach_\text{crit}$. 
    They are calculated using Eq.(\ref{eq:Dpw}).
    Results are shown for different eccentric orbits as a function of the average Mach number. 
    For a circular perturber, $t_\text{steady}=t_\text{crossing}$ regardless of the perturber's velocity.}
    \label{fig:sstimes}
\end{figure}

\subsection{Convergence to a steady dissipation state}

\label{sub:convergence}

Although the acoustic wake never reaches a steady state throughout the entire volume in finite time, the orbit-averaged rates of energy and angular momentum dissipation do. 
Namely, $\langle \dot{E} \rangle$ and $\langle \dot{L}_z \rangle$ become time-independent after a time $t_\text{steady}<\infty$.

Since the FDTD simulations evolve a finite time perturbation, the simulated DF force is given by Eq.~(\ref{eq:wakeforce}) with $h(t')=1$ for $t'>0$. 
Therefore, to estimate $t_\text{steady}$, it is sensible to consider the dissipation rate $\langle \dot E\rangle(t_0)$ of energy 
in the time interval $[t_0,t_0+2\pi/\Omega]$, which is given by
\begin{align}
\label{eq:Edott0}
    \langle \dot E\rangle(t_0) &= (4\pi Gm_p)^2 
    \frac{\bar\rho_g\Omega}{2\pi} \int_{t_0}^{t_0+\frac{2\pi}{\Omega}}\!\!\!\mathrm{d}t \int_\omega\int_{\vk} \int_{-\infty}^{+\infty}\!\! \mathrm{d}t' \\
    &\qquad \times h(t') \frac{(\mathrm{i}\omega)}{k^2}G(\omega,k) \mathrm{e}^{\mathrm{i}\vk\cdot(\vr_p(t)-\vr_p(t'))-\mathrm{i}\omega (t-t')} \nonumber \;.  
\end{align}
Focusing on an ideal fluid and assuming that the perturber is at pericentre passage at $t=0$, 
Appendix~\S\ref{app:math} shows that a steady state of energy dissipation is achieved at the earliest time $t_0$ that satisfies the condition 
\begin{equation}
\label{eq:cplus}
    t_0>c_+(t,t')\equiv\frac{r_p+r_p'}{c_s}-t \quad \forall\, t,t'\in[0,2\pi/\Omega]\,,
\end{equation}
where $r_p = r_p(t)$ and $r_p'=r_p(t')$.
Note that Eq.~(\ref{eq:cplus}) applies to any periodic motion inside an ideal fluid. 
Any perturber moving in periodic motion from $t=0$ onward will reach a steady-state dissipation regime due to DF, after a finite time determined by the maximum 
of the function $c_+(t,t')$ along its trajectory.

To determine the steady-state time $t_\text{steady}$, it is more convenient to parametrize the orbit with the eccentric anomaly and consider
\begin{equation}
    c_+(\xi,\xi')=\frac{a}{c_s}\big(2-e(\cos\xi+\cos\xi')\big)-\frac{1}{\Omega}\big(\xi-e\sin\xi\big)
\end{equation}
This way, Eq.~(\ref{eq:cplus}) defines $t_\text{steady}$ as the least upper bound (supremum) or, equivalently in our context, 
\begin{equation}
\label{eq:tsteady}
    t_\text{steady}=\max\!\big(c_+(\xi,\xi')\big) \quad \forall\, \xi,\xi'\in[0,2\pi] \;.
\end{equation}
For small eccentricities $e\ll 0$, the maximum is reached for $\xi'=\pi$ and $\xi=0$.
For an eccentricity $0<e<1$, the maximum is obtained upon setting $\xi'=\pi$ and $\xi=\xi_\text{max}$, where $\xi_\text{max}$ is either the real root
\begin{equation}
    \label{eq:ximax}
    \xi_\text{max}=\pi-\arcsin\!\left({\frac{1}{e\sqrt{\mach+1}}}\right)-\arctan\!\left({\frac{1}{\mach}}\right)
\end{equation}
of $\frac{\mathrm{d}}{\mathrm{d}\xi}c_+(\xi,\xi')=0$ or $\xi_\text{max}=0$ if there is no such root. 

When $\xi=\xi_\text{max}=0$, steady state is achieved after one sound crossing time $t_\text{crossing}\equiv 2a/c_s$ of the system, as in the circular case \cite{desjacques/etal:2022}. 
We find that this occurs as long as the far focus of the ellipse is closer to the information sphere originating from the pericentre than to the perturber, throughout time.  
To determine when the perturber overtakes the wake, we consider the difference $D$ between two distances: (i) the distance from the perturber located at $\vr_p$ to the far focus, and (ii) the distance from the information sphere -- originating at the pericentre -- to the far focus. Since the latter expands at the sound speed $c_s$, we have
\begin{equation}
\label{eq:Dpw}
    D=\left|\vr_p(t)+2ae\,\xvh\right|+c_st-a(1-e)-2ae \;,
\end{equation}
where $\xvh$ is a unit vector along the semi-major axis ($x$-axis), oriented toward the pericentre. 
A illustration of distances (i) and (ii) is shown in the upper panel of Fig.~\ref{fig:sstimes}.

\begin{figure}
    \centering
    \includegraphics[width=0.48\textwidth]{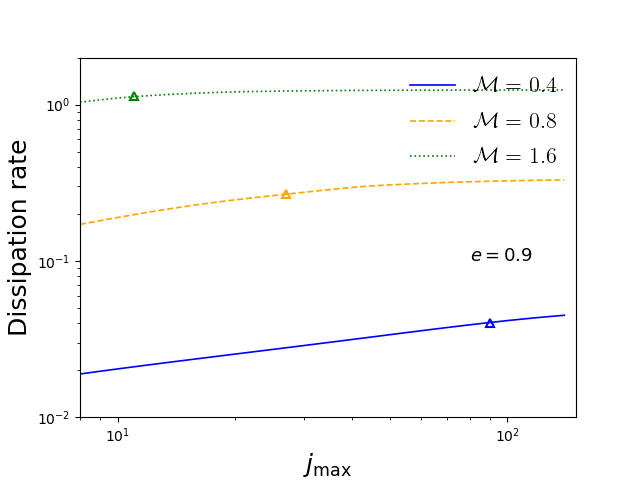}
    \includegraphics[width=0.48\textwidth]{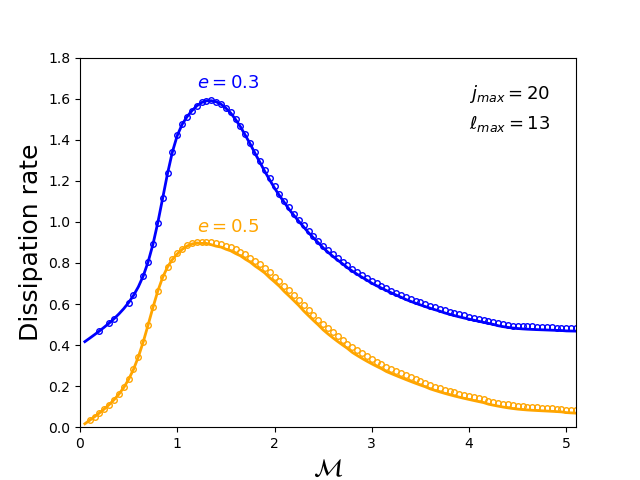} 
    \caption{{\it Upper panel}: Steady dissipation rate (of energy) for $e=0.9$ as a function of $\jmax$ for different values of $\mach$ as indicated in the figure.
    The absissa of the open triangles matches the value of $I_E$ measured in the FDTD simulations. The solid curves show the dependence of the theoretical predictions 
    on $\jmax$ assuming a fixed value of $\lmax=13$.
    {\it Lower panel}: Steady dissipation rate as a function of Mach number for two eccentricities. 
    All data were computed from the theoretical prediction, Eq.~(\ref{eq:Ieccentric}), assuming $\jmax=20$ and $\lmax=13$.
    Open circles represent values obtained via direct numerical integration of the functions $\gjlm(x,e)$.
    The solid curves correspond to results using the "Hansen" series expansion Eq.~(\ref{eq:gjlmhansen}), with $q_\text{max}=70$ (see text for details). 
    The results for $e=0.3$ have been shifted vertically to enhance the clarity of the plot.}
    \label{fig:theoryerrors}
\end{figure}

At the start of the motion, $D=0$ and the perturber moves in the $y$-direction, so $D$ increases. 
If the perturber is fast enough, $D$ can later decrease, and with sufficient speed, it can even become negative. 
At the end, the information sphere reaches the far focus while the perturber never does. Therefore, $D$ must end up positive. 
If $D$ becomes negative, then there exists a minimum value $D_\text{min} < 0$.
Its existence is linked to the real root Eq.~(\ref{eq:ximax}), which determines the maximum of $c_+$. 
When $D_\text{min}<0$ or, equivalently, $\xi_\text{max}>0$, the steady-state time exceeds the crossing time. 
This happens for Mach numbers above a critical value $\mach_\text{crit}$.
For $\mach>\mach_\text{crit}$, we find that $t_\text{steady}$ computed from Eq.~(\ref{eq:tsteady}) increases with $\mach$ and agrees perfectly with the maximal difference between the two distances defined above, i.e. 
\begin{equation}
    \label{eq:tsteadyD}
    t_{\text{steady}}^{\mach>\mach_\text{crit}}=t_{\text{crossing}}-D_{\text{min}}/c_s \;.
\end{equation}
This expression asymptoticaly approaches $(1+e)t_{\text{crossing}}$, which corresponds to the crossing time of twice the apocentre distance.
Appendix \S\ref{app:math} demonstrates that a steady dissipation rate of angular momentum is reached on the same timescale.

The steady-state time $t_\text{steady}$ is shown in the lower panel of Fig.~\ref{fig:sstimes} as a function of $\mach$ for different choices of $e$. 
The breaking points at the critical Mach number $\mach_\text{crit}$, as well as the asymptote $(1+e)t_{\text{crossing}}$, are also indicated in the plot.
For circular motion, both foci coincides and the perturber does not get closer to either of them. Therefore, there is no breaking point. 
Overall, the timescale for convergence to a steady state increases with the eccentricity.

\subsection{Steady dissipation rates}

\label{sub:measurements}

In the FDTD simulations, the work and torque exerted by the acoustic wake on the perturber are computed from the wake gravitational potential and the resulting force field. 
These are obtained by a convolution of the perturbed density with the Newtonian Green’s function kernel \( \frac{G}{|\mathbf{r} - \mathbf{r'}|} \), 
which allows for a direct evaluation in real space.

\begin{figure*}
    \centering
    \includegraphics[width=0.45\textwidth]{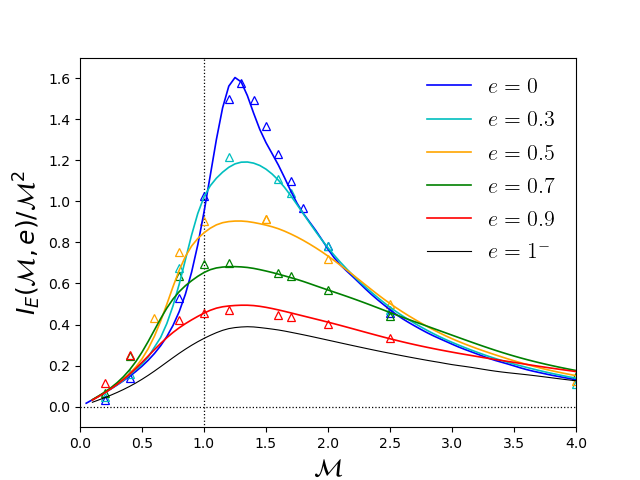}
    \includegraphics[width=0.45\textwidth]{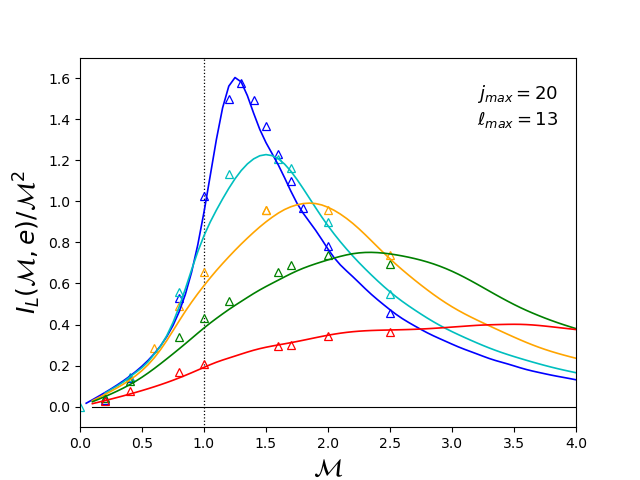}
    \caption{Steady-state friction coefficients $I_E(\mach,e)$ (left panel) and $I_L(\mach,e)$ (right panel) for various eccentricities. 
    The solid curves show the theoretical prediction, Eq.~(\ref{eq:Ieccentric}), assuming the uniform cutoff $\jmax=20$ and $\lmax=13$.
    Open triangles indicate the measurements extracted from a simulation of the linear response acoustic wake (see text for details). 
    }
    \label{fig:EdotLdot}
\end{figure*}

To obtain reliable measurements of the orbit-averaged, steady-state dissipation rates $\langle \dot E\rangle$  and $\langle \dot L_z\rangle$,
the system must reach a steady state within a small fraction of the simulation time (two complete orbital revolutions). 
Since $t_\text{crossing} = \pi^{-1}\mach T$, where $T$ is the orbital period, Fig.~\ref{fig:sstimes} shows that, for eccentricities $e\leq 0.9$, choosing a 
Mach number $\mach\lesssim 2.5$ implies that a steady state is reached on a timescale $t_\text{steady}\lesssim T$. 
In this case, the second, simulated orbital revolution can be used to measure $\langle \dot E\rangle$ and $\langle \dot L_z\rangle$. 

In astrophysical systems, the effective frequency cutoff $\jmax$ is infinite, while the effective multipole cutoff $\lmax<\infty$ and depends primarily 
on the Bondi radius of the perturber, which defines the region within which sound waves do not propagate freely. 
In the FDTD simulations, $\jmax<\infty$ due to the finite time resolution, whereas $\lmax$ is set by the grid resolution. 
The value of $\lmax$ can be determined by a fit to the circular results, which are independent of $\jmax$. We find that $\lmax=12-13$ yields a good fit to 
$I_E(\mach,e=0)$ for $\mach\lesssim 3$. 
By contrast, Eq.~(\ref{eq:jmax}) suggests that the actual value of $\jmax$ varies strongly with $\mach$ and $e$, and could be significantly larger than $\lmax$.
The top panel of Fig.~\ref{fig:theoryerrors}, which illustrates the dependence of $I_E$ on the choice of $\jmax$, confirms this expectation. 
The absissae of the open triangles, whose ordinates match the values of $I_E(\mach,e=0.9)$ measured from the numerical data, indicate that the effective $\jmax$ 
in our simulations lies in the range $\jmax\sim 10 - 100$ and decreases markedly with increasing $\mach$.
We will not pursue this avenue further and, for simplicity, adopt a single value of $\jmax$ in our theoretical predictions.

The top panel of Fig.~\ref{fig:theoryerrors} also demonstrates that, while large effective values of $\jmax\gg 1$ are required in the subsonic regime to accurately 
capture DF on eccentric orbits, the lowest frequency modes are sufficient to recover the correct dissipation rates in the supersonic regime. 
Even though the theoretical predictions presented hereafter rely on a direct numerical integration of $\gjlm(j\mach,e)$, this property could be leveraged
to devise fast estimations of $I_E$ and $I_L$ based on the Hansen series expansion. 
For example, the solid curves in the bottom panel of Fig.~\ref{fig:theoryerrors} are approximations to $I_E$ with $\jmax=20$, $\lmax=13$ and using 
the Hansen series expansion of $\gjlm(j\mach,e)$ restricted to $|j\mach|<30$ (for $e=0.3$) and $|j\mach|<25$ (for $e=0.5$). 
These estimates agree with the results of a direct numerical integration of $\gjlm(j\mach,e)$ (open circles) to within 1.5\% in the range $\mach<5$,
but require significantly less CPU time to 
compute~\footnote{To estimate the steady dissipation rates using the Hansen series expansion, we first pre-compute the Hansen coefficients recursively and store them 
in a cache file. With coefficients pre-computed, the evaluation of $I_E$ and $I_L$ with $\jmax=20$, $\lmax=13$ and our fiducial value of $\sigma_\text{max}=70$ 
(see Appendix \S\ref{app:hansen}) only takes a few milliseconds of CPU time.}.
For $\mach\gg 1$ (not shown in the figure), using only the leading, $\mathcal{O}(\mach^{-3/2})$ term of the stationary phase approximation, Eq.~(\ref{eq:gjlmspa}), 
to evaluate the dissipation rate yields a vanishing friction coefficient at certain values of $\mach$. In this regime, it is therefore necessary to include 
higher order terms in the asymptotic expansion of the spherical Bessel function (only the dominant one is included in Eq.(\ref{eq:spabessel}).  
These contribute corrections of order $\mathcal{O}(\mach^{-5/2})$ and higher.

Fig.~\ref{fig:EdotLdot} displays the corresponding friction coefficients as a function of the Mach number for a range of eccentricities, $0\leq e< 1$.
The open triangles indicate the measurements extracted from the simulations. 
The solid curves show the theoretical predictions from Eq.~(\ref{eq:Ieccentric}), computed using the single cutoff values $\jmax=20$ and $\lmax=13$, as motivated above.
At fixed eccentricity, $\langle\dot E\rangle$ and $\langle\dot{L}_z\rangle$ do not reach their maximum at the same Mach number because the energy dissipation 
depends on both the tangential and radial components $F_\vartheta$ and $F_r$ of the DF force, while $\langle\dot{L}_z\rangle$ depends only on $F_\vartheta$.
At high Mach number, the stationary phase approximation shows that $I_E$ and $I_L$ scale like $\mach^{-2}$. 

\begin{figure*}
    \centering
    \includegraphics[width=0.45\textwidth]{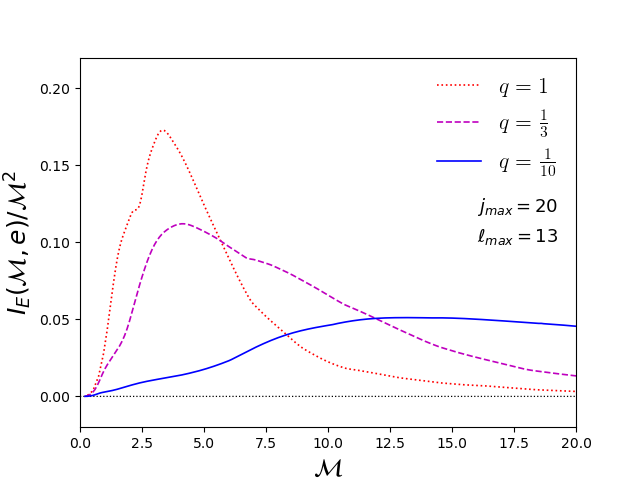}
    \includegraphics[width=0.45\textwidth]{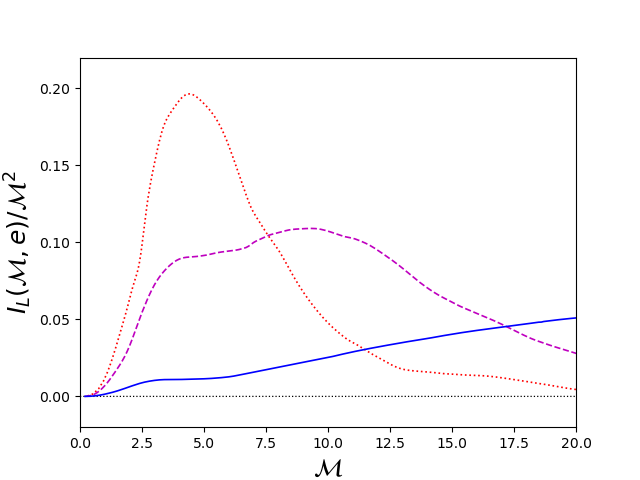}
    \caption{Steady-state friction coefficients for a compact binary of eccentricity $e=0.7$, with mass ratio $q=q_2/q_1=1$, 1/3 and 1/10 as indicated in the figure.
    Results are shown as a function of the characteristic Mach number $\mach = \Omega a/c_s$, where $\Omega$ and $a$ are the binary frequency and semi-major axis.}
    \label{fig:Ibinary}
\end{figure*}

\subsection{Highly eccentric orbits}

For highly eccentric orbits with $|1-e|\ll 1$, a limiting expression for the steady-state dissipation rates can be obtained using 
\begin{equation}
    \lim_{e\to 1^-} \varphi(\xi) = \left\lbrace\begin{array}{cc} 
    \Big. \pi & (0\leq\xi<\pi) \\ \Big. -\pi & (\pi\leq\xi < 2\pi)
    \end{array}\right.
\end{equation}
As a result, $\gjlm(x,e)$ converges to $\lim_{e\to 1^{-}} \gjlm(x,e) = (-\mathrm{i})^m \fjl(x)$, with
\begin{equation}
    \fjl(x) \equiv \frac{1}{\pi}\int_0^\pi\! \mathrm{d}M\,j_\ell\big(x (r_p(M)/a)\big)\cos\!\big(jM\big)
\end{equation}
Substituting $\gjlm(x,e) = (-\mathrm{i})^m \fjl(x)$ in Eq.~(\ref{eq:Ieccentric}) yields the friction coefficients shown as the black solid curves
in Fig.~\ref{fig:EdotLdot}. Note that $I_L$ vanishes in the limit $e\to 1^-$ because the orbital angular momentum also vanishes in this limit,
leaving no angular momentum to be extracted.

\section{Orbital evolution of a compact eccentric binary}
\label{sec:BBH}

Our results can be used to track the secular evolution of a (compact) Keplerian binary characterized by a semi-major axis $a$, eccentricity $e$ 
and companion masses $q_1m_b$, $q_2m_b$ where $q_1\geq q_2$ by definition and $m_b$ is the total mass of the binary. 

\subsection{Secular evolution equations}

The instantaneous rate of change of the orbital elements $a$ and $e$ is given by \cite{burns:1976,murray/dermott:1999}
\begin{align}
\label{eq:evolve}
    \dot a &= -\frac{a}{E}\,\dot E \\
    \dot e &= -\left(\frac{1-e^2}{e}\right) \left(\frac{1}{2E}\,\dot E+\frac{1}{L}\, \dot L\right) \nonumber \;,
\end{align}
where $\dot E$ and $\dot L$ are the instantaneous rate of energy and angular momentum dissipation. 
For a medium with a small enough average density $\bar\rho_g$, the characteristic timescales $\tau_a=|a/\dot a|$ and $\tau_e=|e/\dot e|$ 
over which the orbital elements vary can become much longer than the steady-state time Eq.~(\ref{eq:tsteady}), i.e. $\tau_a,\tau_e\gg t_\text{steady}$. 
In this regime, we can assume that the compact binary evolves through an adiabatic sequence of steady dissipation states, 
with $\dot E\approx \langle \dot E\rangle$ and $\dot L\approx \langle\dot L_z \rangle$ approximately equal to the orbit-averaged dissipation rates of the binary energy and angular momentum. 
Details of their computation can be found in Appendix \S\ref{app:binary}. 
The results can be recast in the form of Eq.~(\ref{eq:coeffriction}), with
\begin{align}
\label{eq:IEILB}
    I_E(\mach,e) &= \mach \sum_{j\ne 0}\sum_{\ell,m} c_{\ell m} \\
    &\quad\times\left(q_2\, g^{(j)}_{\ell m}(q_1 j\mach,e)+(-1)^\ell q_1\, g^{(j)}_{\ell m}(q_2 j\mach,e) \right)^2\nonumber \\
    I_L(\mach,e) &= \mach \sum_{j\ne 0}\sum_{\ell,m} \frac{m}{j}\,c_{\ell m} \nonumber \\
    &\quad\times\left(q_2\, g^{(j)}_{\ell m}(q_1 j\mach,e)+(-1)^\ell q_1\, g^{(j)}_{\ell m}(q_2 j\mach,e) \right)^2\nonumber \;.
\end{align}
Here, $\mach=\Omega a/c_s$ as in the single perturber case.
The functions $\gjlm(q_i j\mach,e)$ can be calculated with the methods discussed in Section \S\ref{sec:Ijlm}. 
The steady-state friction coefficients from Eq.~(\ref{eq:IEILB}) are shown in Fig.~\ref{fig:Ibinary} for several values of the binary mass ratio $q=q_2/q_1$, 
assuming a fixed eccentricity $e=0.7$. 
As the mass ratio decreases, the friction coefficients become less peaked.
These results assume that the self-gravity of the sound wake can be neglected, as in the single perturber case discussed in the previous sections.

\begin{figure}
    \centering
    \includegraphics[width=0.40\textwidth]{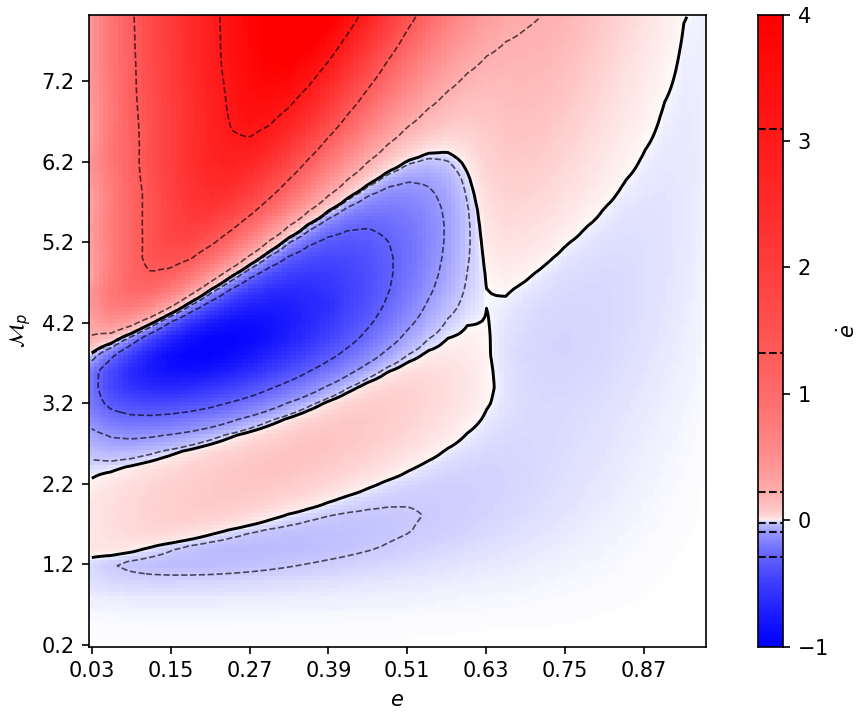}
    \caption{Contours of constant $\dot e$ in the plane $e$--$\mach_p$ for a binary mass ratio $q=1/3$, assuming $\jmax=20$ and $\lmax=13$. 
    The thick black contour indicates the zero level. 
    Regions with $\dot{e}>0$ are shown in red, while those with $\dot{e}<0$ appear in blue.}
    \label{fig:edotplane}
\end{figure}

Concretely, we evolve the binary system according to 
\begin{align}
\label{eq:secular}
    \dot a &= -8\pi G^{1/2}\frac{m_b^{1/2}}{\mu_b} a^{5/2}\,\bar\rho_g \,I_E(\mach,e) \\
    \dot e &= -4\pi G^{1/2}\frac{m_b^{1/2}}{\mu_b} a^{3/2}\,\bar\rho_g \left(\frac{\sqrt{1-e^2}}{e}\right) \nonumber \\
    &\quad \times \Big(\sqrt{1-e^2}I_E(\mach,e)  - I_L(\mach,e)\Big)\nonumber \;,
\end{align}
where $\mu_b$ is the reduced mass, and recall that these really are secular rates of change since the dissipation rates are averaged over one orbital period. 
The quantity $\sqrt{1-e^2} I_E - I_L$, which determines the sign of $\dot e$, can be positive or negative depending on $e$ and $\mach$.
It vanishes for $e=0$ and in the limit $e\to 1^{-}$ regardless of the binary mass ratio (this follows from $\gjlm(x,e)\to (-\mathrm{i})^m f_{j\ell}(x)$ and $c_{\ell,-m}=c_{\ell,m}$).
For illustration, Fig.~\ref{fig:edotplane} shows the rate of change $\dot{e}$ (in unit of $4\pi G^{1/2}\frac{m_b^{1/2}}{\mu_b} a^{3/2}\,\bar\rho_g$) for binary mass ratio $q=1/3$. 
Contour levels are shown in the plane $e$ - $\mach_p$, where 
\begin{equation}
    \mach_p=\frac{\mach}{(1+q)}\sqrt{\frac{1+e}{1-e}}
\end{equation}
is the Mach number of the lighter companion at pericentre. 
For eccentricities $e\lesssim 0.6$, there are two separate regions in which $\dot{e}<0$ and orbits circularize. 
For large values of $\mach_p$, $\dot{e}$ is positive throughout most of the $e$ - $\mach_p$ plane, suggesting that all the orbits eventually become highly eccentric.

\begin{figure*}
    \centering
    \includegraphics[width=0.45\textwidth]{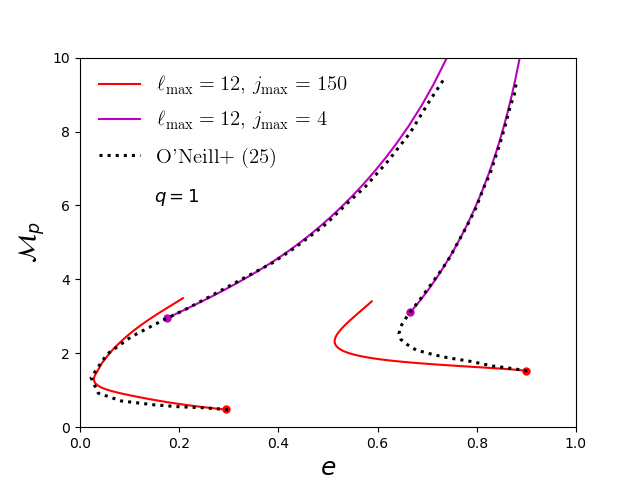}
    \includegraphics[width=0.45\textwidth]{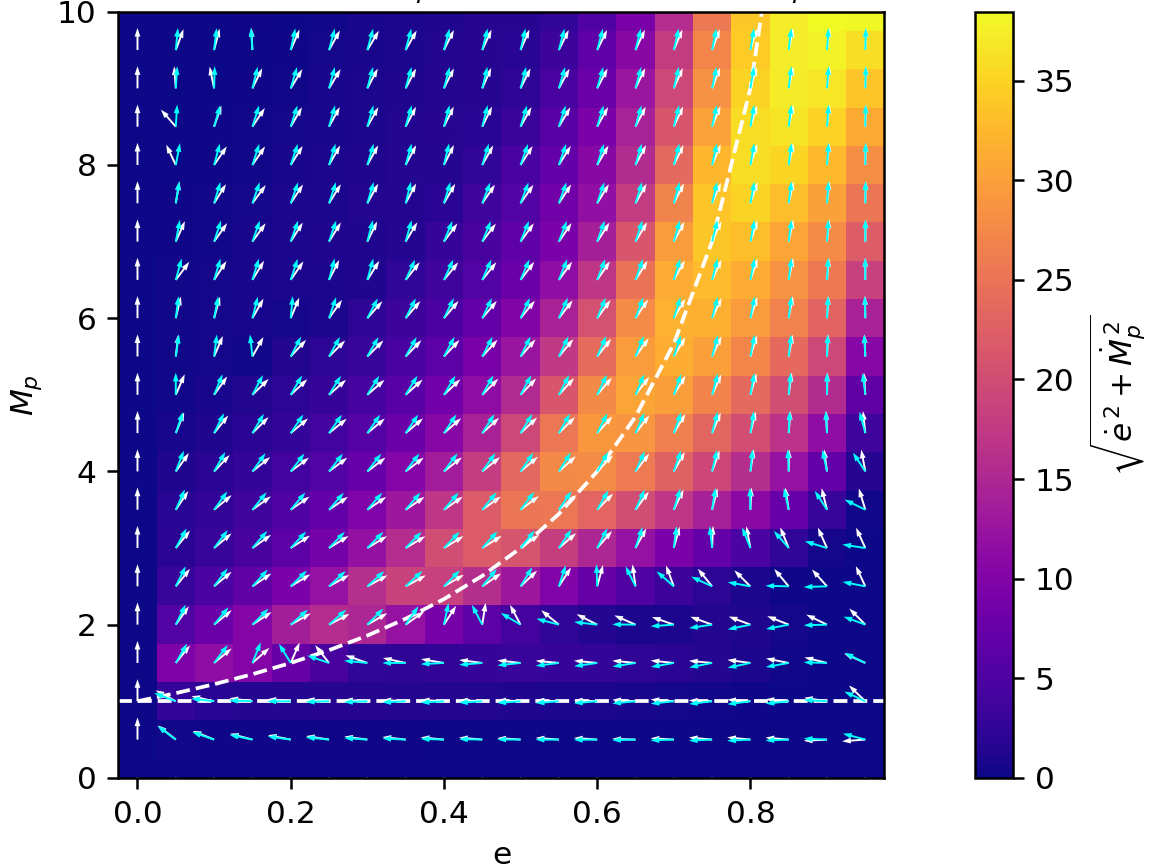}
    \caption{{\it Left panel}: Comparison between the orbital evolution of an equal-mass binary ($q=1$) predicted by the numerical approach of \cite{oneill/etal:2024} 
    (thick dotted curves) and our analytical formulation (thin solid curves). The latter assume a fixed value of $\lmax=12$ and two different choices of $\jmax$ 
    as indicated in the figure. Results are shown in the $e$ - $\mach_p$ plane for two different orbits. 
    {\it Right panel}: Magnitude and direction of the vector field $(\dot e,\dot{\mach}_p)$, which determines the flow in the  $e$ - $\mach_p$ plane.
    The results of \cite{oneill/etal:2024} are shown as the white arrows. Our results shown as the cyan arrows assume $\ell_\text{max}=12$ and $j_\text{max}=20$. 
    The dashed lines divide the grid into a subsonic region where $\mach_p<1$, a supersonic region where $\mach_p>\frac{1+e}{1-e}$, and a transonic region in between.}
    \label{fig:compare}
\end{figure*}

\subsection{Comparison to previous work}

Numerical simulations of the dynamical friction acting on a compact binary system have been presented in \cite{kim/etal:2008,oneill/etal:2024} using a semi-numerical model based on 
linear response theory. In particular, Ref.~\cite{oneill/etal:2024} analyzed how the drag produced by an ideal fluid affects the secular evolution of equal-mass, compact Keplerian orbits.

In the left panel of Fig.~\ref{fig:compare}, the thick dotted curves show two orbital evolution tracks extracted from Fig.~15 of \cite{oneill/etal:2024}, plotted in the $e$--$\mach_p$ plane. 
These integral curves were computed from a discrete sample of the vector field $(\dot e,\dot\mach_p)$ (see \cite{oneill/etal:2024} for details).
The thin solid curves are solutions to the secular equations~(\ref{eq:secular}), with initial conditions marked by filled circles and chosen to lie on the evolutionary tracks of 
\cite{oneill/etal:2024}.
Our orbital evolution tracks assume $\lmax=12$, as motivated by the comparison in \cite{oneill/etal:2024} with the drag force measured by \cite{kim/etal:2008} for an equal-mass, 
circular compact binary (This drag force is reproduced by the analytical solution of \cite{desjacques/etal:2022} when $\lmax\approx 12$).
In contrast, we have considered two widely separated values of $\jmax$ -- 4 and 150 -- to emphasize that the effective $\jmax$ of the simulated tracks computed in \cite{oneill/etal:2024} 
varies significantly across the orbital evolution. 

We find that, at fixed $\lmax=12$, the effective $\jmax$ required to match their results is significantly larger during the early, 
subsonic stage of the orbital evolution. This is consistent with the trend observed in our simulation of a single eccentric perturber, see Fig.~\ref{fig:theoryerrors}.
To understand why the high eccentricity case is off relative to \cite{oneill/etal:2024} in the early stage of its evolution, note that in the simulations of \cite{oneill/etal:2024}
distances from the point masses shorter than $r_\text{min}=0.1a$ do not contribute to the force (like in \cite{kim/etal:2008}).
As discussed in \cite{desjacques/etal:2022}, this is analogous to an azimuthal resolution of $\frac{\pi R}{r_\text{min}}$ -- where $R$ is the binary separation -- which corresponds
to an effective multipole cutoff $\lmax<\infty$. 
However, while for a circular binary $R=a$ and $\lmax$ is thus constant throughout the orbit, $a(1-e)\leq R\leq a (1+e)$ for eccentric motions. 
This implies that $\lmax$ will change with the orbital phase, unless $r_\text{min}$ is varied during the simulation such that $\lmax$ remains constant. 
We have ignored this complication in the comparison with \cite{oneill/etal:2024}.

\begin{figure*}
    \centering
    \includegraphics[width=0.45\textwidth]{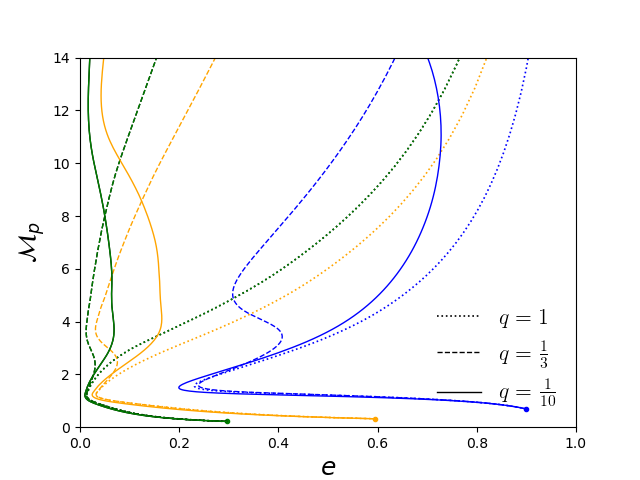}
    \includegraphics[width=0.45\textwidth]{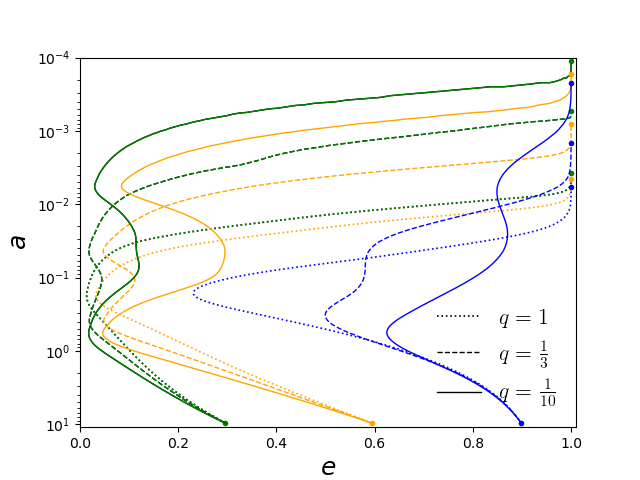}
    \caption{Binary orbital evolution in the plane $e$ - $\mach_p$ (left panel) and $e$ - $a$ (right panel) for various initial conditions and binary mass ratios $q=1$, 1/3 and 1/10. 
    Time flows from bottom to top (toward large $\mach_p$ and small $a$). All these theoretical predictions assume $\jmax=\lmax=20$.}
    \label{fig:secular}
\end{figure*}

The vector field $(\dot e,\dot\mach_p)$ shown in the right panel of Fig.~\ref{fig:compare} illustrates the flow of orbital trajectories in the $e$--$\mach_p$ plane for equal-mass
binaries, as in Fig.12 of \cite{O’Neill_2024}. 
The white arrows show the results of \cite{O’Neill_2024}, sampled on a regular two-dimensional grid. These are compared to our predictions, shown as cyan arrows, which are obtained 
from Eq.~(\ref{eq:secular}) using the relation between $\dot{\mach}_p$ and the time derivatives $\dot{a}$ and $\dot{e}$ of the orbital parameters. 
We have assumed $\lmax=12$ and, for simplicity, a fixed $\jmax=20$.
Note that, while $\dot\mach>0$ always (since the binary loses energy), $\dot{\mach}_p$ can be positive or negative depending on $\dot e$.
We observe that $\dot e$ is negative for strictly subsonic motion, positive for strictly supersonic motion and changes sign for transonic motion, in agreement with the findings of 
\cite{oneill/etal:2024}. 
We find slight deviations from the measurements of \cite{O’Neill_2024} presumably caused by the $\mach$-dependence of the effective frequency cutoff $\jmax$ in their simulations. 

To conclude, note that when $\lmax\lesssim 15$, numerical estimates of the friction coefficient $I_L$ can yield negative values in the regime $\mach\gg 1$ due to the sensitivity of 
the alternating sum to numerical errors. 
Therefore, it is important to use effective multipole cutoffs of at least $\lmax\sim 20$ to ensure that the orbital trajectories are well-behaved deep in the supersonic regime. 
For $\lmax=20$, we find that orbits converge at Mach numbers $\mach_p\gtrsim 2$ if $\jmax\gtrsim 20$. 

\subsection{Sensitivity to the binary mass ratio}

Fig.~\ref{fig:secular} shows the dependence of the secular binary evolution on the binary mass ratio $q$. 
Results are shown for $q=1$, 1/3 and 1/10, assuming three different initial conditions with initial eccentricities $e_i=0.3$, 0.6 and 0.9.
Furthermore, we have assumed $\jmax=\lmax=20$ to ensure convergence at high Mach numbers (see above).
The left panel focuses on the early time evolution of the binary orbits, which are shown in the $e$--$\mach_p$ plane to facilitate comparison with Fig.~\ref{fig:compare}.

Dynamical friction reaches a local maximum whenever the pericentre velocity of a binary component is comparable to $c_s$.
For equal-mass binaries, eccentric orbits approaching the sound barrier from below tend to circularize, as the strongest drag is experienced at pericentre.
Once they become supersonic, however, the drag is strongest at apocentre and causes the orbits to become more eccentric 
(see \cite{oneill/etal:2024,buehler/etal:2024} for a detailed discussion of this effect).
As seen in the left panel of Fig.~\ref{fig:secular}, a minimum eccentricity is reached when $\mach_p\sim 1 - 2$ when the orbit becomes transonic.

For unequal mass binaries, the compact objects do not reach the sound speed at the same time. 
Since the Mach number of the more massive companion is $q\mach_p<\mach_p$, the lighter object becomes transonic before the heavier one for $q<1$. 
As a result, local eccentricity minima occur twice along an orbit, first for the lighter companion at $\mach_p\sim 1$ and then for the massive companion at $\mach_p\sim q^{-1}$ 
(since the pericentre velocity of the massive companion is $q\mach_p$). 
Fig.~\ref{fig:secular} shows that the timing of these events sensitively depends on the value of $q$ and $e$.
For $q=1/3$, a second eccentricity minimum is visible at $\mach_p\sim 3 - 6$ (consistent with the location of the blue region with $\dot{e}<0$ in Fig.~\ref{fig:edotplane})
whereas, for $q=1/10$, a second eccentricity minimum occurs at $\mach_p\gtrsim 10$. 

These local eccentricity minima are more apparent for all mass ratios considered in the right panel of Fig.~\ref{fig:secular}, which shows the long-term orbital evolution in the $e$ - $a$ plane. 
The filled circles at the end of the orbits mark the point at which $1-e\leq 10^{-6}$ is reached for the first time. 
All orbits eventually attain high eccentricity, but this occurs at smaller values of $a$ -- that is, after a larger amount of energy has been dissipated via dynamical friction -- 
the smaller the binary mass ratio. This delay is due to the second local minimum in the evolution of the orbital eccentricity.

\section{Conclusions}

\label{sec:conclusions}

We have used a multipole–frequency decomposition to calculate the shape of the density wake and the dynamical friction (DF) induced by a point-like perturber on an eccentric orbit 
within a homogeneous, fluid-like medium. Our results are valid within the framework of linear response theory and in the regime where the self-gravity of the acoustic wake can be 
neglected.

We have shown how the orbit-averaged, steady dissipation rates of orbital energy and angular momentum, $\langle\dot E\rangle$ and $\langle \dot L_z\rangle$, can be evaluated 
for arbitrary medium Green's functions and eccentricities. 
We have derived explicit expressions for $\langle\dot E\rangle$ and $\langle \dot L_z\rangle$, as well as the steady-state acoustic wake, in the specific case of an ideal fluid 
where energy and angular momentum are carried away from the perturber exclusively through sound modes. 
These predictions have been validated with numerical simulations of the linear density response produced by a perturber switched on at pericentre at $t=0$. 

For such a finite-time (step-like) perturbation, the orbit-averaged rates of orbital energy and angular momentum dissipation reach a steady state after a finite steady-state time 
$t_\text{steady}$, which we have computed for generic eccentricity and Mach number. 
At fixed eccentricity, the steady-state time is bounded by $2a/c_s\leq t_\text{steady}\leq 2a(1+e)/c_s$.
To gain physical intuition, note that the steady-state acoustic wake appears at the origin after a time $t=a(1+e)/c_s$ following the activation of the perturber at $t=0$.
Since the radius of the steady-state density region grows at the sound speed, $c_st$, the lower bound $2a/c_s$ corresponds to the time required for the steady-state wake 
to reach the pericentre, while the upper bound $2a(1+e)/c_s$ corresponds to the time required to reach the apocentre.
Were the fluid's self-gravity turned on, these bounds would remain valid for perturbation wavelengths much smaller than the Jeans scale.

We have applied our results to model the secular evolution of unequal mass, compact Keplerian binaries embedded in a homogeneous ideal fluid (which can be regarded as a zeroth order 
approximation to a realistic "gaseous" medium). We have assumed small medium densities, such that the orbital elements vary on a timescale much longer than $t_\text{steady}$. 
In this approximation, compact binaries evolve through an adiabatic sequence of steady dissipation states. 
Dynamical friction reaches a local maximum whenever the pericentre velocity of a binary component is comparable to $c_s$.
For equal-mass binaries, our evolution tracks recover the early circularization and the late-time eccentricity growth discussed in \cite{oneill/etal:2024,buehler/etal:2024}.
For unequal-mass binaries, we find that the smaller the mass ratio $q<1$, the longer the system spends at low eccentricities (for initial eccentricities $e_i\lesssim 0.6$).
This is due to the fact that the binary companions do not become transonic at the same time, so that local eccentricity minima occur twice along the orbit. 
As a result, the late-time orbital eccentricity growth is significantly delayed in systems with $q\ll 1$.

In the harmonic approach used here, practical evaluations of the dissipation rates depend on the frequency and multipole cutoffs, $\jmax$ and $\lmax$. 
In linear response theory, dynamical friction (DF) is divergent in the limit $\lmax\to\infty$ for supersonic circular orbits \cite{desjacques/etal:2022}, 
and we expect a similar divergence to persist for eccentric orbits.
Therefore, while in a fully nonlinear approach (beyond the scope of this paper) the magnitude of DF would be fully determined by the underlying physics (fluid equation of state,
accretion onto the perturber etc.) once $\lmax$ is sufficiently large, in linear response theory it remains sensitive to the exact value of $\lmax$.
As a rule of thumb, $\lmax$ should increase with decreasing $r_B/a$, where $r_B$ is the Bondi radius of the perturber (within which there are no freely propagating sound waves), 
and $a$ is the semi-major axis.
We have considered several numerical techniques to accelerate the evaluation of the dissipation rates, which can be computationally expensive when the frequency and/or multipole cutoffs 
are large. These numerical methods should help exploring a wider range of compact binary systems -- including extreme binary mass ratios, dissipation via gravitational wave emission, 
wave-like dark matter backgrounds -- within a relatively short computational time. Our harmonic approach can also be extended to collisionless media such as stars and dark matter. 
We leave this for future work. 

\acknowledgments

We thank Frans van Die, Aleksey Generosov, Adi Nusser and Hagai Perets for useful discussions. 
V.D.~acknowledges support from the Israel Science Foundation (grant no.~2562/20).
This work was supported by a Leverhulme Trust International Professorship Grant (no.~LIP-2020-014). 
The work of Y.B.G.~was partly supported by a Simons Investigator Award to A.A.~Schekochihin.

\appendix

\section{Derivation of $\langle \dot E\rangle$ for ideal fluid}

\label{app:math}

To evaluate Eq.~(\ref{eq:Me}), we exploit the periodicity of the unperturbed path $\vr_p(t)$ to divide the infinite integral over $t'$ into orbital periods,
\begin{align}
\label{eq:Edotproc}
    \langle\dot E\rangle &= (4\pi Gm_p)^2 \bar\rho_g\frac{\Omega}{2\pi}  \int_0^{\frac{2\pi}{\Omega}}\!\mathrm{d}t \int_\omega\int_{\vk} \sum_n \int_{\frac{2\pi n}{\Omega}}^{\frac{2\pi(n+1)}{\Omega}}\! \mathrm{d}t' \nonumber \\ &\quad \times\frac{\mathrm{i}\omega}{k^2}\,
  G(\omega,k)\,\mathrm{e}^{\mathrm{i}\vk\cdot(\vr_p(t)-\vr_p(t'))-\mathrm{i}\omega (t-t')} \\
  &= (4\pi Gm_p)^2 \bar\rho_g\frac{\Omega}{2\pi}  \int_0^{\frac{2\pi}{\Omega}}\!\mathrm{d}t \int_\omega\int_{\vk} \int_0^{\frac{2\pi}{\Omega}}\! \mathrm{d}t' \nonumber \\
  &\quad \times\frac{\mathrm{i}\omega}{k^2}\,
  G(\omega,k)\,\mathrm{e}^{\mathrm{i}\vk\cdot(\vr_p(t)-\vr_p(t'))-\mathrm{i}\omega (t-t')}\sum_n \mathrm{e}^{\mathrm{i}\frac{2\pi n}{\Omega}\omega} \nonumber \;.
\end{align}
Using the relation
\begin{equation}
\label{eq:expsum}
    \sum_n \mathrm{e}^{\mathrm{i}\frac{2\pi n}{\Omega}\omega}=\Omega\sum_j\delta_D(\omega - j\Omega)
\end{equation}
and the Rayleigh expansion
\begin{equation}
    \label{eq:expexpansion}
    \mathrm{e}^{\mathrm{i}\vk\cdot\vr}=4\pi\sum_{\ell=0}^{\infty}\sum_{m=-\ell}^{+\ell}\mathrm{i}^\ell j_\ell(kr)Y_\ell^m(\hat{\vk})Y_\ell^{m*}(\hat{\vr}) \;,
\end{equation}
we arrive at Eq.~(\ref{eq:Edotkintegral}) after taking advantage of the orthogonality of the spherical harmonics to eliminate the integral over $\hat{\vk}$.

For a finite time perturbation characterized by $h(t)=1$ for $t>0$ and zero otherwise, the rate of energy dissipation can be computed using the representation
\begin{equation}
\label{eq:stepfunction}
h(t')=-\frac{1}{2\pi \mathrm{i}}\lim_{\varepsilon\rightarrow 0^+}\int_{-\infty}^{\infty}\!\mathrm{d}\omega'\,\frac{\mathrm{e}^{-\mathrm{i}\omega't'}}{\omega'+\mathrm{i}\varepsilon} \;.
\end{equation}
Starting from the general expression Eq.~(\ref{eq:Edott0}) for the dissipation rate $\langle \dot E\rangle(t_0)$ of energy in the time interval $[t_0,t_0+2\pi/\Omega]$, 
and repeating the steps outlined in the derivation of Eq.~(\ref{eq:Edotproc}), we obtain
\begin{align}
    \langle \dot E\rangle(t_0)&= -\frac{(4\pi Gm_p)^2}{2\pi \mathrm{i}} \bar\rho_g \frac{\Omega^2}{2\pi}
    \int_{t_0}^{t_0+\frac{2\pi}{\Omega}}\!\mathrm{d}t \int_\omega\int_{\vk} \int_0^{\frac{2\pi}{\Omega}}\! \mathrm{d}t'\,\nonumber\\
    &\times \int_{-\infty}^{\infty}\!\mathrm{d}\omega'\frac{\mathrm{e}^{-\mathrm{i}\omega't'} }{\omega'+\mathrm{i}0^+}\frac{\mathrm{i}\omega}{k^2}\,
    \frac{\mathrm{e}^{\mathrm{i}\vk\cdot(\vr_p(t)-\vr_p(t'))-\mathrm{i}\omega (t-t')}}{c_s^2 k^2 - (\omega + \mathrm{i}\varepsilon)^2}\nonumber\\
    &\times \bigg. \sum_j\delta_D(\omega - \omega' - j\Omega) 
\end{align}
upon replacing $G(\omega,k)$ by the retarded Green's function of a homogeneous and isotropic ideal fluid.
Here and henceforth, we use $0^+$ as a shorthand for $\lim_{\varepsilon\to 0^+}$.

The Dirac delta function eliminates the $\omega$-integral and sets $\omega=\omega'+j\Omega$. 
Substituting the Rayleigh expansion, we obtain the following result:
\begin{widetext}
\begin{align}
    \langle \dot E\rangle(t_0) &= -(4\pi Gm_p)^2 \bar\rho_g \frac{\Omega^2}{4\pi^4}
    \sum_{j=-\infty}^\infty\sum_{\ell=0}^\infty\sum_{m=-\ell}^\ell 
    \int_{t_0}^{t_0+\frac{2\pi}{\Omega}}\!\mathrm{d}t \int_0^{\frac{2\pi}{\Omega}}\!\mathrm{d}t'\int_{-\infty}^{\infty}\! \mathrm{d}\omega'\,\frac{\mathrm{e}^{-\mathrm{i}\omega't'}}{\omega'+\mathrm{i}0^+} 
    \int_0^\infty\!\mathrm{d}k\,\mathrm{e}^{-\mathrm{i}\,(\omega'+j\Omega)(t-t')} \nonumber \\
    &\qquad \times \frac{(\omega'+j\Omega)j_\ell(k r_p)\,j_\ell(k r_p')}{c_s^2 k^2 - (\omega'+j\Omega + \mathrm{i}\varepsilon)^2}\,Y_\ell^m(\rvh_p)Y_\ell^{m*}(\rvh_p') \;.
\end{align}
To carry out the $k$-integral, the following integral identity can be used \cite[e.g.][]{desjacques/etal:2022}
\begin{equation}
    \label{eq:Sintegral}
    S_{l,l-1}^{\alpha,\beta}(x)=\lim_{\varepsilon\rightarrow0^+}\int_0^\infty \mathrm{d}z\frac{zj_l(\alpha z)j_{l-1}(\beta z)}{z^2-(x+\mathrm{i}\varepsilon)^2}
    =\left\lbrace \begin{array}{ll}
        \frac{\pi}{2}[\mathrm{i}j_{l-1}(\beta x)h_l^{(1)}(\alpha x)-\left(\frac{\beta^{l-1}}{\alpha^{l+1}}\right)\frac{1}{x^2}] & (\alpha>\beta) \\
        \frac{\mathrm{i}\pi}{2}j_l(\alpha x)h_{l-1}^{(1)}(\beta x) & (\alpha<\beta) \end{array}\right. \;.
\end{equation}
On performing the change of variable $t\to t-t_0$, we arrive at
\begin{align}
\label{eq:EdotS}
    \langle \dot E\rangle(t_0) &= -(4\pi Gm_p)^2 \frac{\bar{\rho_g}\Omega^2}{4\pi^4 c_s^2}
    \sum_{j=-\infty}^\infty\sum_{\ell=0}^\infty\sum_{m=-\ell}^\ell \int_{-\infty}^{\infty}\! \mathrm{d}\omega' \int_0^{\frac{2\pi}{\Omega}}\!\mathrm{d}t \int_0^{\frac{2\pi}{\Omega}}\!\mathrm{d}t'\,
    \frac{(\omega'+j\Omega)\,\mathrm{e}^{-\mathrm{i}\,\omega'(t_0+t)+\mathrm{i}j\Omega (t'-t_0-t)}}{(2l+1)(\omega'+\mathrm{i}0^+)}\,\nonumber \\
    &\qquad \times r_p\left[S_{l,l-1}^{r_p',r_p}\!\left(\frac{\omega'+j\Omega}{c_s}\right)
    + S_{l+1,l}^{r_p,r_p'}\!\left(\frac{\omega'+j\Omega}{c_s}\right)\right]\,Y_\ell^m(\rvh_p)Y_\ell^{m*}(\rvh_p')\;,
\end{align}
\end{widetext}
where $r_p$, $\rvh_p$ are now evaluated at a time $t_0+t$.
Using the recursion relations of spherical Bessel functions, we can recast part of the integrand into
\begin{align}
    \label{eq:twoSfunctions}
    r_p\Big(S_{l,l-1}^{r_p',r_p}(x)&+S_{l+1,l}^{r_p,r_p'}(x)\Big)=\frac{\mathrm{i}\pi}{2x} \\
    &\times\left[\big(2\ell+1\big)j_\ell(r_< x)\, h_\ell^{(1)}\!(r_> x) + \mathrm{i}\frac{r_<^\ell}{r_>^{\ell+1}}\frac{1}{x}\right]\nonumber
\end{align}
with the shorthand notations $r_<=\min(r_p,r_p')$ and $r_>=\max(r_p,r_p')$, while $x=\frac{\omega'+j\Omega}{c_s}$ is assumed.
Importantly, the Laurent expansion of Eq.~(\ref{eq:twoSfunctions}) around $x=0$ is analytic. Therefore, in the $\omega'$-plane there is only one (simple) pole at $\omega'=0$.

The $\omega'$-integral can be solved with Cauchy's Residue theorem. To determine where to draw the contours, we express the spherical Bessel function as 
a superposition of outgoing and incoming waves, i.e. $j_\ell(x) = \frac{1}{2}(h_\ell^{(1)}\!(x)+h_\ell^{(2)}\!(x))$ where $h_l^{(1)}(z)\sim \mathrm{e}^{\mathrm{i}z}$ and $h_l^{(2)}(z)\sim \mathrm{e}^{-\mathrm{i}z}$. 
For large values of $t_0$, the coefficient of $\omega'$ in the exponent (including the ingoing and outgoing waves) is negative and thus all contours must be drawn in the lower half of the $\omega'$-plane and enclose the pole at the origin. 
Since the dissipation rate is real, only the real part in the square brackets of Eq.~(\ref{eq:twoSfunctions}) can contribute to the final expression, which reads
\begin{align}
    \langle \dot E\rangle(t_0) &= -(4\pi Gm_p)^2 \frac{\bar\rho_g}{c_s}\left(\frac{\Omega}{2\pi}\right)^2
    \sum_{j=-\infty}^\infty\sum_{\ell=0}^\infty\sum_{m=-\ell}^\ell  \\
    &\qquad\times \int_0^{\frac{2\pi}{\Omega}}\!\mathrm{d}t \int_0^{\frac{2\pi}{\Omega}}\!\mathrm{d}t'\,\mathrm{e}^{\mathrm{i}j\Omega (t'-t_0-t)}\,\nonumber \\
    &\qquad \times j_\ell\left(j\mach\frac{r_<}{a}\right)j_\ell\left(j\mach\frac{r_>}{a}\right)\,Y_\ell^m(\rvh_p)Y_\ell^{m*}(\rvh_p')\nonumber \;. 
\end{align}
On replacing $r_>$, $r_<$ by $r_p'$, $r_p$, exploiting the periodicity of the $t$-integrand to shift $t$ by $t_0$ and noticing that the $t$ and $t'$ integrals can be 
re-written in terms of $I_{j\ell m}(j\mach,e)$ and $I_{j\ell m}^*(j\mach,e)$, we precisely get the steady-state result with a friction coefficient $I_E$ given by 
Eq.~(\ref{eq:Ieccentric}).

The earliest time $t_0$ after which the coefficient of $\omega'$ is negative for all $t',t\in[0,2\pi]$ determines the time $t_\text{steady}$ 
at which a steady dissipation regime is achieved. 
Since the decomposition of the spherical Bessel function into incoming and outgoing waves involves two exponentials with coefficients $c_\pm-t_0$, 
where $c_\pm=\frac{r_>\pm r_<}{c_s}-t$. $c_+>c_-$, it is sufficient to demand that 
\begin{equation}
    t_0>c_+(t,t')\equiv\frac{r_<+r_>}{c_s}-t 
\end{equation}
be satisfied for all the times $t$, $t'$ in the interval $[0,2\pi/\Omega]$.
This motivates the definition (\ref{eq:tsteady}) of $t_\text{steady}$.

A similar calculation can be carried out for the dissipation rate $\langle \dot L_z\rangle(t_0)$ of angular momentum in the time interval $[t_0,t_0+2\pi/\Omega]$. 
The expression of $\langle \dot L_z\rangle(t_0)$ is identical to Eq.~(\ref{eq:EdotS}) except that the multiplicative  factor of $(\omega'+j\Omega)$ is replaced by $m$.
Since there is again the same simple pole at $\omega'=0$ solely, a steady dissipation state of angular momentum is reached on the same timescale $t_\text{steady}$.

\section{Wake density for ideal fluid}
\label{app:wake}

In an ideal fluid, the sound wake density in the steady-state limit can be computed straightforwardly from Eq.~(\ref{eq:deltarho}). 
On setting $h(t')=1$ and following the steps outlined in Eqs.~(\ref{eq:Edotproc}) -- (\ref{eq:expexpansion}) yields 
\begin{align}
    \label{eq:delrhokintegral}
    \delw(t,\vr) &=
    \frac{4}{\pi}\bar\rho_g Gm_p\Omega\sum_{j}\sum_{lm}\int_{0}^{\frac{2\pi}{\Omega}}\mathrm{d}t'\int_{0}^{\infty}\mathrm{d}k\\ 
    &\times \frac{k^{2}\mathrm{e}^{-\mathrm{i}j\Omega(t-t')}}{c_{s}^{2}k^{2}-(j\Omega+\mathrm{i}\varepsilon)^{2}}
    j_\ell(kr)j_\ell(kr_p')Y_{\ell}^{m}(\rvh)Y_{\ell}^{m*}(\rvh_p')
    \nonumber
\end{align}
for the density perturbation. 
For $j\ne 0$, the $k$-integral can be performed with the Residue theorem, which gives the identity
\begin{align}
    \int_{0}^{\infty}\!\mathrm{d}k&\,\frac{kJ_{\ell+\frac{1}{2}}\left(kr\right)J_{\ell+\frac{1}{2}}\left(kr_{p}^{\prime}\right)}{k^{2}-(\frac{j\Omega}{c_{s}}+\mathrm{i}\varepsilon)^{2}}\\
    &=\frac{\mathrm{i}\pi}{2} J_{\ell+\frac{1}{2}}\!\left(j\mathcal{M}\frac{r_{<}}{a}\right)\, H_{\ell+\frac{1}{2}}^{(1)}\!\left(j\mathcal{M}\frac{r_{>}}{a}\right)\nonumber
\end{align}
where $r_<=\min(r,r_p')$ and $r_>=\max(r,r_p')$.
For $j=0$, it can be solved using the identity
\begin{equation}
    \int_0^\infty \!\mathrm{d}k\, j_{\ell}(kr)\, j_{\ell}(kr_p^\prime)=\frac{\pi}{2(2\ell+1)}\frac{r_<^{\ell}}{r_>^{\ell+1}}\;.
\end{equation}
Inserting this result into Eq.~(\ref{eq:delrhokintegral}) gives the steady-state expression (\ref{eq:delrhotint}).
It is important to note that a steady-state density wake is reached only asymptotically, as sound waves must propagate through the entire (infinite) volume.

For a finite time (step-like) perturbation, we use the representation of $h(t)$ in Eq.~(\ref{eq:stepfunction}) and find, after some algebra, 
\begin{align}
    \label{eq:delrhofinite}
    \delw(t,\vr) &= - \frac{\mathrm{i}\bar\rho_g \mach}{2\pi \mathrm{i}} \frac{r_B}{a}\sum_{j}\sum_{lm}\int_{0}^{\frac{2\pi}{\Omega}}\mathrm{d}t'\mathrm{e}^{-\mathrm{i}j\Omega(t-t')}\\
    &\times \int_{-\infty}^{\infty}\mathrm{d}\omega'\left(j\Omega+\omega'\right)\frac{\mathrm{e}^{-\mathrm{i}\omega't}}{\omega'+\mathrm{i}0^+}
    j_\ell\!\left(\frac{j\Omega+\omega'}{c_{s}}r_{<}\right) \nonumber \\
    &\times h_\ell^{(1)}\!\left(\frac{j\Omega+\omega'}{c_{s}}r_{>}\right)\,Y_{\ell}^{m}(\rvh)\,Y_{\ell}^{m*}(\rvh_p')\nonumber \;.
\end{align}
The integral over $\omega'$ can be computed analogously to Eq.~(\ref{eq:EdotS}).
Decomposing the spherical Bessel function into an incoming and an outgoing wave, the coefficients multiplying $\omega'$ in the exponents of $\exp(\mathrm{i}c_\pm \omega')$ 
are $c_\pm=\frac{r_>\pm r_<}{c_s}-t$. 

For a point either sufficiently close to the origin $\vr=0$, or at a sufficiently late time such that 
$t>\frac{r_>+r_<}{c_s}=\frac{r+r_p'}{c_s}$
for all $t'\in [0,2\pi]$,
the contours must be drawn in the lower half of the $\omega'$-plane and thus enclose the pole at $\omega'=0$.
As a consequence, the steady-state solution is reached, i.e. 
\begin{equation}
\delw\!\left(t>\frac{r+\rapo}{c_s},\vr\right)=\delw^\text{steady}(t,\vr) \;.
\end{equation}
We see that the steady-state density region is a sphere of radius $r<c_s t - \rapo$. 
This holds for any periodic motion inside an ideal fluid.
For a point far enough from the origin such that $t<\frac{r_>-r_<}{c_s}$ for all $t'\in[0,2\pi]$, 
the contours are drawn in the upper half-plane, from which it follows that $\delw(t,\vr)=0$. 
Physically, this indicates that the information sphere has not yet reached this point. 

\begin{figure}
    \centering
    \includegraphics[width=0.45\textwidth]{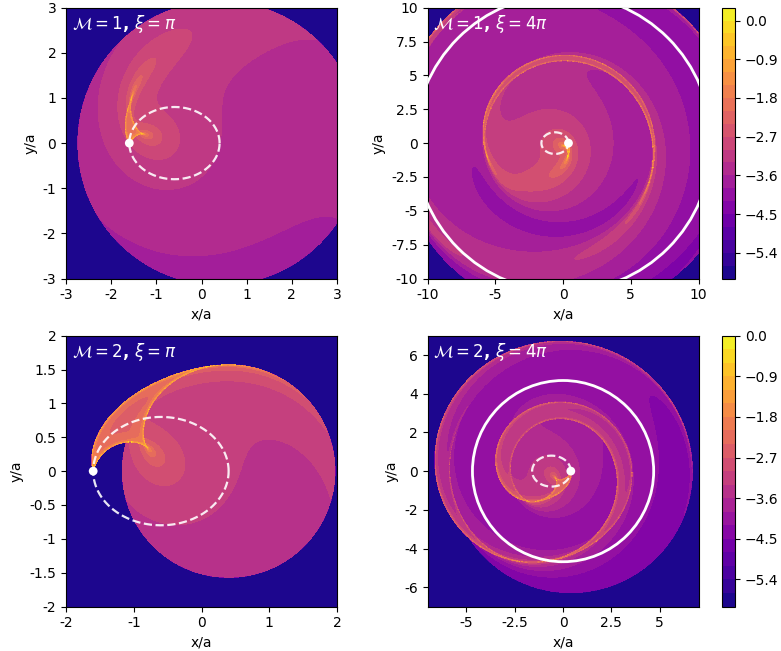}
    \caption{Wake overdensity $\log_{10}(\delrho/\bar\rho_g)$ calculated using Eq. (\ref{eq:delrhoOstriker}) for a eccentric orbit with $(e,\mach)=(0.6,1)$ (upper panels) and $(e,\mach)=(0.6,2)$ (lower panels). The left panels display the acoustic wake after half a cycle, while the right panels show the acoustic wake after two cycles. The perturber's position is represented as a filled (white) symbol, and its orbit as a dashed (white) ellipse. The large (white) circles enclose the steady state region.}
    \label{fig:densities}
\end{figure}

To validate these results, note that for an ideal fluid $\delw(t,\vr)$ can be also calculated directly using the Green's function in real space, 
\begin{equation}
    G(\tau,\vr) = \frac{1}{r}\delta_D\left(\tau-\frac{r}{c_s}\right)\;.
\end{equation}
As demonstrated in \cite{ostiker:1999}, $\delw(t,\vr)$ is given by
\begin{align}
\label{eq:delrhoOstriker}
    \delw(t,\vr)&=\frac{GM\bar\rho_g}{c_{s}^{2}}\int_{-\infty}^{\infty}\mathrm{d}t' h(t')\frac{\delta_{D}\left(t-t'-\frac{1}{c_{s}}\left|\vr-\vr_p'\right|\right)}{\left|\vr-\vr_p'\right|}
    \nonumber \\
    &=\frac{GM\bar\rho_g}{c_s^2}\sum_{t'_i}\frac{h(t'_i)}{\left|1+\frac{\partial|\vr_p(t')-\vr|}{c_s\partial t'}\big|_{t'=t'_i}\right||\vr_i-\vr|}
\end{align}
where $\vr_p'=\vr_p(t')$, $\vr_i=\vr_p(t'_i)$ and $t'_i$ are the roots of the delta function.
Substituting an eccentric Keplerian orbit as done in \cite{O’Neill_2024} enables us to compute the density wakes shown in Fig. \ref{fig:densities} for a transonic ($\mach=1$) and
a supersonic ($\mach=2$) orbit of eccentricity $e=0.6$.

We have checked that, apart from numerical errors, the density wake reaches the steady-state solution inside the information sphere (indicated by the large white circle). 
By subtracting the density wakes of trajectories that differ by an integer number of orbital cycles, we found that the difference within the steady-state region remains below 
a relative error of $\frac{\delrho_1-\delrho_2}{\delrho_{1/2}} < 10^{-6}$. 
We have also observed that the actual steady-state region is slightly larger and not perfectly spherical. 
We believe this region may correspond to $D$ in Eq.(\ref{eq:Dpw}). The wake size, in reality, is smaller than the spherical estimate derived from Eq.(\ref{eq:delrhofinite}) and 
instead appears to trace the envelope of all information spheres emitted along the perturber's trajectory, as shown in Fig.~\ref{fig:densities}.

\section{Hansen coefficients}

\label{app:hansen}

The Hansen coefficients \cite{Hansen1855} are defined by the generating function \cite{plummer:1918}
\begin{equation}
\label{eq:HansenDefinition}
    \left(\frac{r_p}{a}\right)^n\, \mathrm{e}^{\mathrm{i}m\varphi} = \sum_{s=-\infty}^{+\infty} X_s^{n,m}\!(e)\, \mathrm{e}^{\mathrm{i} s M} \;.
\end{equation}
Tables of Hansen coefficients can be found in \cite{laskar/boue:2010}.
For practical computations, the Hansen coefficients are conveniently expressed in terms of the Newcomb operators $X^{n,m}_{\rho,\sigma}$ introduced in \cite{newcomb:1895}, 
and can be efficiently evaluated using the recursion relations
\begin{equation}
\label{eq:HansenNewcomb}
    X^{n,m}_k=e^{|k-m|}\sum_{\sigma=0}^{\infty}X^{n,m}_{\sigma+\alpha,\sigma+\beta}\, e^{2\sigma}
\end{equation}
given in \cite{hughes:1981}. Here, $\alpha=\max(0,k-m)$ and $\beta=\max(0,m-k)$. 
In practice, the sum is truncated at a maximum value $\sigma_\text{max}$, which makes the accuracy of the Hansen coefficients dependent on the eccentricity. 
From Eq. (\ref{eq:HansenNewcomb}) and the properties of Newcomb coefficients, we find $X^{n,m}_k=X^{n,-m}_{-k}$. This identity is used throughout our calculations.
Our fiducial choice of $\sigma_\text{max}=70$ implies that $X^{n,m}_k(e)$ is computed up to $\mathcal{O}(e^{140})$. 
In principle, the higher the eccentricity, the higher the order to which $X^{n,m}_k(e)$ should be computed.

\section{Stationary phase approximation}

\label{app:phase}

The method of stationary phase can be used to approximate the integral of a product of a smooth function $f(t)$ times the exponent $\mathrm{e}^{\mathrm{i}x\psi(t)}$ of a real-valued function
$\psi(t)$ in the limit $x\rightarrow\infty$. At leading order, we have
\begin{align}
    \label{eq:spa}
    \int\!\mathrm{d}t\, f(t)\mathrm{e}^{\mathrm{i}x\psi(t)}&\approx\sum_{t_i}\sqrt{\frac{2\pi}{x|\psi^{\prime\prime}|}}e^{s\left\{ \psi^{\prime\prime}\right\} \mathrm{i}\pi/4}\mathrm{e}^{\mathrm{i}x\psi(t_{0})}f(t_{0})
\end{align}
where the sum is taken over all the stationary points $t_i$ satisfying $\nabla f=0$, 
while $s\{\psi''\}$ is a shorthand for the sign of the second derivative of $\psi(t)$ at the point $t_i$. 
Higher order corrections are of the form
\begin{equation}
    \sum_{t_i}\mathrm{e}^{\mathrm{i}x\psi(t_{0})}\sum_{n=1}^{\infty}\frac{\sqrt{2\pi}\left(2n-1\right)!!\,e^{s\left\{ \psi^{\prime\prime}\right\} \mathrm{i}(n+\frac{1}{2})\pi/2}}{\left(x|\psi^{\prime\prime}|\right)^{n+\frac{1}{2}}\left(2n\right)!}f^{(2n)}(t_{0})
\end{equation}
and scale as $x^{-n-1/2}$ (with $n\geq 1$). These corrections are for the stationary phase approximation. 
In the case of Eq.~(\ref{eq:gjlm}), we use the asymptotic approximation of $j_\ell(z)$ for large $z\gg\ell$.
With the change of variable $M\rightarrow\xi$ (from the mean to the eccentric anomaly), the function $\gjlm(x,e)$ becomes
\begin{align}
\label{eq:spabessel}
    \lim_{x\rightarrow\infty}\gjlm(x,e)&= \frac{1}{4\pi \mathrm{i} x}\int_{0}^{2\pi}\!\mathrm{d}\xi\,\bigg(\mathrm{e}^{\mathrm{i}\left(x\left(1-e\cos\xi\right)-\frac{l}{2}\pi\right)} \\
    &\quad -\mathrm{e}^{-\mathrm{i}\left(x\left(1-e\cos\xi\right)-\frac{l}{2}\pi\right)}\bigg)\,\mathrm{e}^{\mathrm{i}\left(m\varphi-j\left(\xi-e\sin\xi\right)\right)}\nonumber
\end{align}
for large arguments $x\gg\ell$. Using Eq.~(\ref{eq:spa}) with the stationary points $\xi_i=0$ and $\pi$ yields the asymptotic approximation given in Eq.~(\ref{eq:gjlmspa}).

\section{Orbit-averaged dissipation rates for a compact eccentric binary}

\label{app:binary}

For a compact eccentric binary, the orbit-averaged dissipation rate of energy and angular momentum due to dynamical friction can be computed from the relations
\begin{align}
    \langle\dot E\rangle &= \frac{1}{T}\int_\gamma\! \mathrm{d}t\,\dot\vr\cdot\big(q_1\vvf_2 - q_2\vvf_1\big) \\
    \langle\dot L_z\rangle &= \frac{1}{T}\int_\gamma\!\mathrm{d}t\, \vr\times\big(q_1\vvf_2 - q_2\vvf_1\big) \nonumber \;.
\end{align}
Here, $\vr(t)=\vr_2(t)-\vr_1(t)$ is the binary separation vector and $\gamma$ describes one orbital revolution.
Furthermore, $\vr_i(t)$ and $\vvf_i$ are the position and the force exerted by the total density wake on the $i$th companion of mass $q_im_b$, 
where $q_1+q_2=1$ and $m_b$ is the total binary mass.
For instance, the force acting on the second companion is
\begin{align}
    \vvf_2 &= - \big(q_2 m_b\big)\, \grad_\vx \tilde\phi(\vx,t)\Big\lvert_{\vx=\vr_2(t)} \\
    &= -\big(q_2 m_b\big)\,\grad_{\vr(t)}\tilde\phi\big(\vr_\text{cm}(t)+q_1\vr(t),t\big) \, \frac{1}{q_1} \nonumber 
\end{align}
since $\vr_2(t)=\vr_\text{cm}(t) + q_1\vr(t)$. Furthermore,
\begin{widetext}
\begin{equation}
    \tilde\phi(\vx,t) = -(4\pi G)^2m_b\bar\rho_g \int_\omega\int_{\vk} \int_{-\infty}^{+\infty}\! \mathrm{d}t'\,\frac{h(t')}{k^2}\,G(\omega,k)\,
    \Big[q_1\, \mathrm{e}^{\mathrm{i}\vk\cdot(\vx-\vr_1(t'))} + q_2\, \mathrm{e}^{\mathrm{i}\vk\cdot(\vx-\vr_2(t'))} \Big]\, \mathrm{e}^{-\mathrm{i}\omega (t-t')}
\end{equation}
is the gravitational potential produced by the combined acoustic wakes trailing the compact objects. 
Ignoring small, time-dependent perturbations in the position of the binary center-of-mass $\vr_\text{cm}$, we can write
\begin{equation}
    \label{eq:tildephi2}
    \tilde\phi(\vr_\text{cm}+q_1\vr,t) \approx -(4\pi G)^2 m_b\bar\rho_g\int_\omega\int_{\vk} \int_{-\infty}^{+\infty}\! \mathrm{d}t'\,\frac{h(t')}{k^2}\,G(\omega,k)\,
    \Big[q_1\, \mathrm{e}^{\mathrm{i}\vk\cdot(q_1\vr+q_2\vr')} + q_2\, \mathrm{e}^{\mathrm{i}\vk\cdot(q_1\vr-q_1\vr')} \Big]\, \mathrm{e}^{-\mathrm{i}\omega (t-t')} \;,
\end{equation}
where $\vr=\vr(t)$ and $\vr'=\vr(t')$.
Likewise, the gravitational force acting on the first companion reads $\vvf_1 = q_2^{-1}(q_1 m_b)\,\grad_{\vr(t)}\tilde\phi\big(\vr_\text{cm}(t)-q_2\vr(t),t\big)$, with
\begin{align}
    \label{eq:tildephi1}
    \tilde\phi(\vr_\text{cm}-q_2\vr,t) \approx -(4\pi G)^2m_b\bar\rho_g \int_\omega\int_{\vk} \int_{-\infty}^{+\infty}\! \mathrm{d}t'\,\frac{h(t')}{k^2}\,G(\omega,k)\,
    \Big[q_1\, \mathrm{e}^{\mathrm{i}\vk\cdot(-q_2\vr+q_2\vr')} + q_2\, \mathrm{e}^{\mathrm{i}\vk\cdot(-q_2\vr-q_1\vr')} \Big]\, \mathrm{e}^{-\mathrm{i}\omega (t-t')} \;,
\end{align}
\end{widetext}
Exploiting the fact that the right-hand side of equations~(\ref{eq:tildephi2}) and (\ref{eq:tildephi1}) depends only on $\vr$, we can use Stokes' theorem to express
the orbit-averaged rate of energy dissipation as
\begin{align}
    \langle\dot E\rangle &= \frac{m_b\Omega}{2\pi} 
    \int_\gamma\!\mathrm{d}t\,\frac{\partial}{\partial t}\Big(q_2\tilde\phi(\vr_\text{cm}+q_1\vr,t) \\
    &\qquad +q_1 \tilde\phi(\vr_\text{cm}-q_2\vr,t)\Big) \nonumber \\
    &= -8\pi\big(Gm_b\big)^2 \bar\rho_g\Omega \int_0^{\frac{2\pi}{\Omega}}\!\mathrm{d}t \int_{-\infty}^{+\infty}\! \mathrm{d}t'\,\left(\frac{\partial}{\partial t}\right) \nonumber \\
    &\qquad\times\int_\omega\int_{\vk} \,\frac{G(\omega,k)}{k^2}\, \Delta(q_1,q_2;\vk,\vr,\vr')\,\mathrm{e}^{-\mathrm{i}\omega (t-t')} \nonumber \;.
\end{align}
after setting $h(t')\equiv 1$. In addition, we get
\begin{align}
    \langle\dot{L}_z\rangle &= -\frac{m_b\Omega}{2\pi} 
    \int_\gamma\!\mathrm{d}t\,\vr\times\grad_\vr\Big(q_2\tilde\phi(\vr_\text{cm}+q_1\vr,t) \\
    &\qquad +q_1\tilde\phi(\vr_\text{cm}-q_2\vr,t)\Big) \nonumber \\
    &= -8\pi \big(Gm_b\big)^2 \bar\rho_g \int_0^{\frac{2\pi}{\Omega}}\!\mathrm{d}t \int_{-\infty}^{+\infty}\! \mathrm{d}t'\,\left(L_z\right) \nonumber \\
    &\qquad\times\int_\omega\int_{\vk} \,\frac{G(\omega,k)}{k^2}\, \Delta(q_1,q_2;\vk,\vr,\vr')\,\mathrm{e}^{-\mathrm{i}\omega (t-t')} \nonumber 
\end{align}
for the rate of angular momentum dissipation. Note that the function 
\begin{align}
    \Delta(q_1,q_2;\vk,\vr,\vr') &= \Big(q_2 \mathrm{e}^{\mathrm{i}q_1\vk\cdot\vr}+q_1 \mathrm{e}^{-\mathrm{i}q_2\vk\cdot\vr}\Big) \\
    &\qquad \times \Big(q_2 \mathrm{e}^{-\mathrm{i}q_1\vk\cdot\vr'}+q_1 \mathrm{e}^{\mathrm{i}q_2\vk\cdot\vr'}\Big) \nonumber
\end{align}
does not depend explicitly on time. 
Applying now the same methodology developed for a single perturber (see Section \S\ref{sub:steady}) leads to the steady-state dissipation
rates given in Eq.~(\ref{eq:IEILB}) for a compact eccentric binary.

\bibliography{references, references_fromSISpaper}

\end{document}